\documentclass[12pt, draftclsnofoot, onecolumn]{IEEEtran}
\IEEEoverridecommandlockouts

\usepackage{amsmath,amssymb,amsfonts}
\usepackage{graphicx} 
\usepackage{subfigure}
\usepackage{bm}
\usepackage{cite}
\usepackage{flushend}
\usepackage{amsfonts}
\usepackage{stfloats} 

\usepackage{booktabs}
\usepackage{tabularx}
\usepackage{amssymb}
\usepackage{graphicx}
\usepackage{threeparttable}

\usepackage{makecell}

\usepackage[ruled]{algorithm2e}
\usepackage{algorithmic}

\usepackage{extarrows}

\usepackage{bbding}

\usepackage{tabularx}

\usepackage{cuted} 

\usepackage{extarrows}

\usepackage{color}

\usepackage{graphicx}

\usepackage[bookmarks=false]{hyperref}

\ifCLASSINFOpdf

\else

\fi

\begin{document}

\title{Integrated Sensing and Communications for V2I Networks: Dynamic Predictive Beamforming for Extended Vehicle Targets}

\author{Zhen~Du,~\IEEEmembership{Graduate Student Member, IEEE},
Fan~Liu,~\IEEEmembership{Member, IEEE},
Weijie~Yuan,~\IEEEmembership{Member, IEEE},
Christos~Masouros,~\IEEEmembership{Senior Member, IEEE},
Zenghui~Zhang,~\IEEEmembership{Senior Member, IEEE},
Shuqiang~Xia,~\IEEEmembership{} and  
Giuseppe~Caire,~\IEEEmembership{Fellow, IEEE}
\thanks{This paper has been partly submitted to 2022 IEEE International Conference on Acoustics, Speech and Signal Processing (ICASSP), 22-27 May 2022, Singapore. (\textit{Corresponding author: Fan Liu)}}
\thanks{Zhen Du is with the School of Electronic Information and Electrical Engineering, Shanghai Jiao Tong University, Shanghai 200240, China, and also with the Department of Electrical and Electronic Engineering, Southern University of Science and Technology, Shenzhen 518055, China. (email: duzhen@sjtu.edu.cn).}
\thanks{Fan Liu and Weijie Yuan are with the Department of Electrical and Electronic Engineering, Southern University of Science and Technology, Shenzhen 518055, China (email: liuf6@sustech.edu.cn; yuanwj@sustech.edu.cn).}
\thanks{Christos Masouros is with the Department of Electronic and Electrical
Engineering, University College London, London WC1E 7JE, U.K. (email: chris.masouros@ieee.org)}
\thanks{Zenghui Zhang is with the School of Electronic Information and Electrical Engineering, Shanghai Jiao Tong University, Shanghai 200240, China (email: zenghui.zhang@sjtu.edu.cn).}
\thanks{Shuqiang Xia is with ZTE Corporation, Shenzhen, China, and also with State Key Laboratory of Mobile Network and Mobile Multimedia Technology, Shenzhen, China (email: xia.shuqiang@zte.com.cn).}
\thanks{Giuseppe Caire is with the Electrical Engineering and Computer Science Department, Technische Universit\"at Berlin, 10587 Berlin, Germany (email: caire@tuberlin.de).}
}

\maketitle

\begin{abstract}
We investigate sensing-assisted beamforming for vehicle-to-infrastructure (V2I) communication by exploiting integrated sensing and communication (ISAC) functionalities at the roadside unit (RSU). The RSU deploys a massive multi-input-multi-output (mMIMO) array at mmWave. The pencil-sharp mMIMO beams and fine range-resolution implicate that the point-target assumption is impractical, as the vehicle's geometry becomes essential. Therefore, the communication receiver (CR) may never lie in the beam, even when the vehicle is accurately tracked. To tackle this problem, we consider the extended target with two novel schemes. For the first scheme, the beamwidth is adjusted in real-time to cover the entire vehicle, followed by an extended Kalman filter to predict and track the position of CR according to resolved scatterers. An upgraded scheme is proposed by splitting each transmission block into two stages. The first stage is exploited for ISAC with a wide beam. 
Based on the sensed results at the first stage, the second stage is dedicated to communication with a pencil-sharp beam, yielding significant communication improvements. We reveal the inherent tradeoff between the two stages in terms of their durations, and develop an optimal allocation strategy that maximizes the average achievable rate. Finally, simulations verify the superiorities of proposed schemes over state-of-the-art methods.
\end{abstract}


%
\IEEEpeerreviewmaketitle

\section{Introduction} \label{sec1}
The intelligent transportation system (ITS) has been playing an increasingly important role in building smart cities, which requires the information interaction among vehicles, pedestrians, obstacles, and infrastures \cite{dimitrakopoulos2010,2012Intelligent,zhang2011data}. To achieve such a goal, vehicle-to-everthing (V2X) communication has become a key technique in ITS, which is currently supported by two potential technologies: dedicated short-range communications (DSRC) \cite{kenney2011dedicated} and Cellular V2X (C-V2X) \cite{gonzalez2018tvt,2016Millimeter}. However, DSRC can not meet the demand of the high data traffic, particularly in high vehicle density scenarios \cite{abboud2016}. Besides, conventional C-V2X schemes such as LTE-V2X can only provide localization services at the accuracy of $10m$ and latency of $1s$ \cite{xiao2020overview}. This is unable to meet the critical demand of future V2X networks, requiring high-accuracy localization services on the order of a centimeter and latency on the order of a millisecond in high-mobility scenarios \cite{wymeersch2017}.

Fortunately, the state-of-the-art V2X communication is supported by the fifth-generation (5G) communication techniques, and in particular, the mMIMO and mmWave technologies \cite{wymeersch2017}. 
To be specific, by exploiting the mMIMO antenna array pencil-sharp beams can be generated, to compensate for the mmWave propagation path loss, providing large array gain while improving the angular resolution for vehicle sensing. Moreover, both communication rate and ranging resolution are significantly improved, thanks to the large bandwidth at the mmWave band. Critically, the sparsity of mmWave channel enables the much fewer Non Line-of-Sight (NLoS) components relative to the sub-6 GHz band, which is in favor of the vehicle localization \cite{wymeersch2017}. 
Featured with the above attributes, it is sought to equip the V2X network with both sensing and communication capabilities, such that the vehicles can better interact with surroundings to realize driving safety.
In consideration of all these perspectives, recent efforts have been taken towards the feasibility of integrated sensing and communication (ISAC) in V2X networks \cite{gonzalez2016,liu2020TWC,yuan2020,liu2020tutorialpart1,liu2020tutorialpart2,liu2020tutorial}.

\begin{figure}[!t]
	\centering
	\includegraphics[width=3.75in]{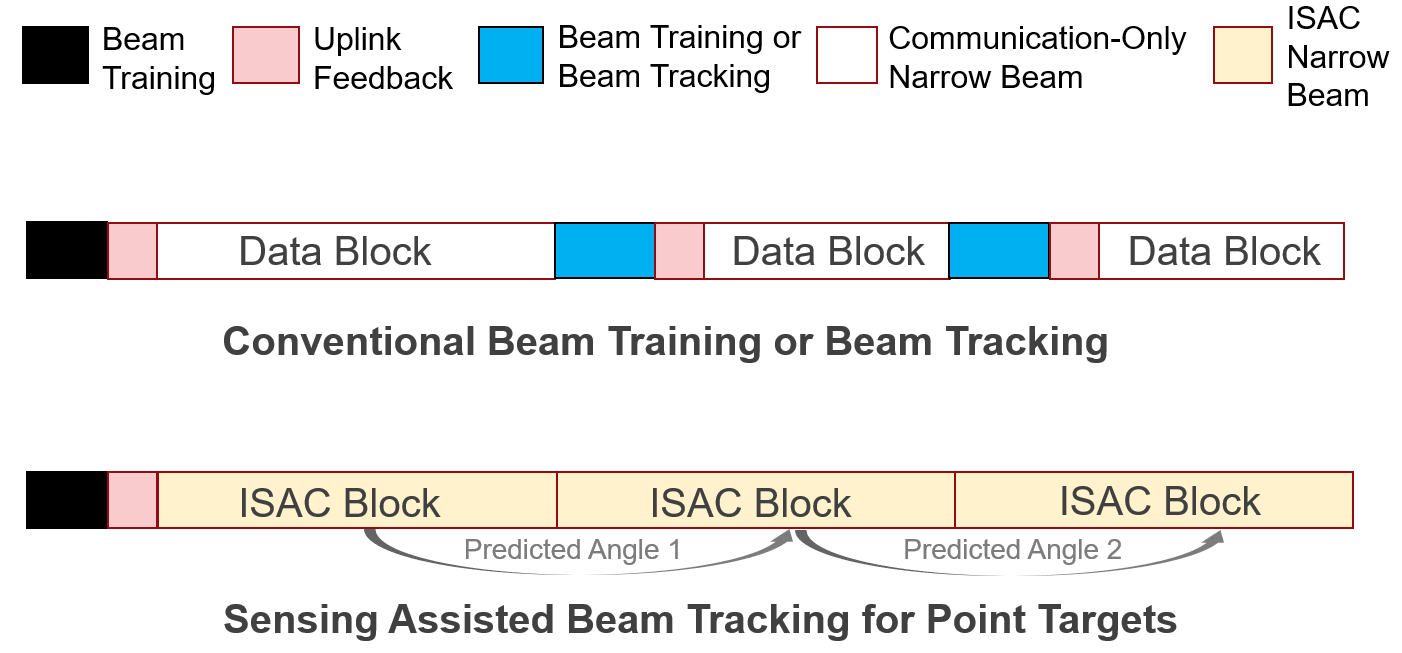}\\
	\caption{Frame structures of the conventional beam traning/tracking method and sensing-assisted  beam tracking for point targets method.}\label{fig1}
\end{figure}

More relevant to this work, a mmWave vehicle-to-infrastructure (V2I) mmWave communication scenario is considered in \cite{gonzalez2016}, where the roadside unit (RSU) is equipped with a dedicated radar sensor to aid the communication beamforming. Such a scheme leads to the more precise V2I beam alignment while reducing the beam training overheads, whereas at the cost of extra hardwares. To further exploit the performance gain of sensing-assisted communication in V2X networks, \cite{liu2020TWC} proposed an ISAC approach to use the mmWave ISAC signaling for predictive beamforming in high mobility V2I scenarios. The echo of the downlink signal reflected from the vehicle is collected for estimating the vehicle's state parameters (distance, velocity, AoA, radar cross section, etc) with matched-filtering \cite{richards2014} at the current time instant. Then an EKF approach is exploited to predict the position of vehilce so as to accurately align the beam.
The benefit of this methodology is that the predictive beamforming is achieved without extra devices, i.e., a single signal is exploited to realize two functionalities of communications and sensing. Further, \cite{yuan2020} approached the same objective from a Bayesian perspective and utilized the message passing algorithm, which provides an enhanced tracking performance over the extended Kalman filtering (EKF) algorithm used in \cite{liu2020TWC}.
Overall, in contrast to the conventional beam training \cite{wang2009,zhang2019} or beam tracking \cite{zhu2017auxiliary,liu2019ekf} which are built upon the pure communication protocols, the advantages of the sensing-assisted beam tracking with the ISAC signaling can be summarized as follows:
\begin{itemize}
	\item[$\bullet$] No dedicated downlink pilots, i.e. a small subset of the data frame, and uplink feedback are required \cite{liu2021integrated}. Conventional beam training or beam tracking demands the RSU to transmit pilots ahead of each data block in the dowinlink channel. Then the vehicle can estimate the state information and feed it back to the RSU in the uplink channel. Sensing-assisted beam tracking can totally release such high overheads burden, as shown in Fig. \ref{fig1};
	\item[$\bullet$] Significant matched-filtering gain \cite{liu2021integrated}. 
	For conventional beam training or beam tracking, only a very limited number of pilots are used. On the contrary, the ISAC approach exploits the whole data frame for both sensing and communication. As such, and despite the echo power attenuation caused by the round-trip path loss, the matched filtering gain of sensing is much more significant. As a result, the state estimation can be more precise due to the higher receive signal-to-noise ratio (SNR);
	\item[$\bullet$] The extra ranging information on top of the angle estimation provides additional degrees of freedom (DoFs) for more accurately localizing the vehicle. This forms the basis of this paper where a dynamic beamwidth is determined by the angle and the distance simultaneously. 
\end{itemize}

The aforementioned works are dedicated to handling beam tracking of point-like vehicles. In practical scenarios, however, the vehicle as a target may be extended in both range and angle domains, and the precise position of the CR in the vehicle is key in beam tracking. Consequently, the approaches discussed in \cite{liu2020TWC,yuan2020,liu2020tutorialpart1,liu2020tutorialpart2,liu2020tutorial} may suffer from significant performance loss, and even fail to provide a reliable V2I link. 
As an example, a 128-antenna mMIMO array is able to synthesize a pencil-sharp beam, with a  beamwidth at the order of $1^\circ$-$2^\circ$. This suggests that, when treating it as a point-like target, the vehicle is unlikely to be entirely illuminated by the narrow beam, especially when it approaches the illuminator, i.e., the RSU in our case. Owing to this reason, even if the vehicle can be accurately tracked, the position of communication receiver (CR) deployed on the vehicle may be beyond the effective region of beamwidth, resulting in beam misalignment and link outages.
Moreover, the 5G mmWave signaling occupies a bandwidth up to the order of GHz, indicating that the vehicle may be distributed in multiple range cells (e.g., for a bandwidth of $B=500$MHz, the range resolution is $c/(2B)=3e^8/(2\times5e^8)=30cm$ \cite{richards2014}, where $c$ is the speed of light). In others words, even if the CR is in the effective beamwidth at present, 
the point-like modeling does not allow the high-resolution sensing to locate the correct range cell that the CR lies in. Therefore, the wrong range cell as the input to the beam tracker may lead to considerable tracking errors.

\begin{figure}[!t]
	\centering
	\includegraphics[width=3.5in]{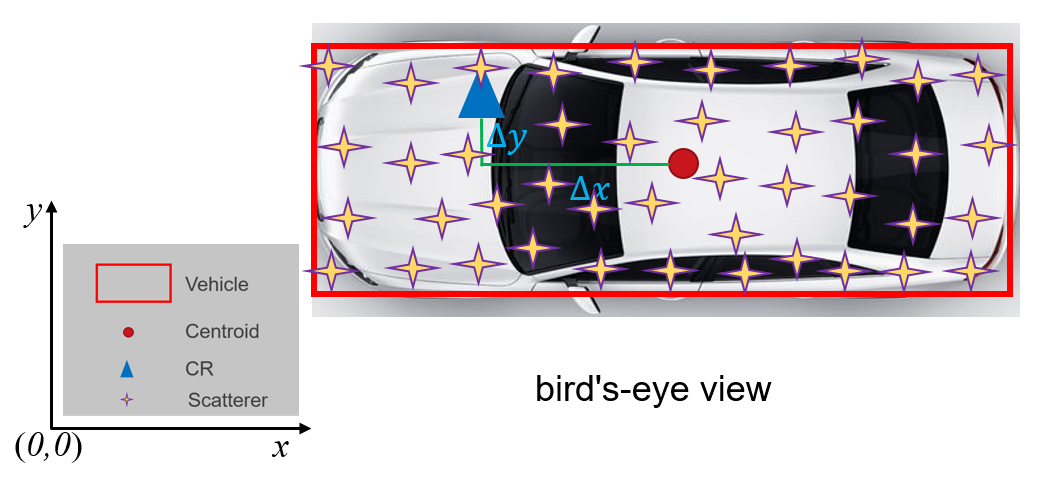}\\
	\caption{A schematic diagram of the extended target vehicle with multiple uniformly distributed scatterers.}\label{30.png}
\end{figure}

To overcome the above challenges, we propose the sensing-assisted predictive beam tracking in V2I networks by modeling the vehicle as an extended target. The major novelties of this article can be summarized in accordance with the proposed two schemes:
\begin{itemize}
	\item[$\bullet$] \textbf{ISAC-based predictive beam tracking with dynamic beam (ISAC-DB):}
	Inspired by the geometry of the extended vehicle target with multiple resolvable scatterers as shown in Fig. \ref{30.png}, the coordinates of the CR can be inferred according to measured state vectors of all scatterers. Subsequently, the beam alignment can be performed using a tailored EKF algorithm. In contrast to \cite{liu2020TWC} where a pencil-sharp beam with fixed beamwidth is generated to track a point-like vehicle target, in this work the beamwidth must be adjusted in real-time in accordance with the predicted distance and the angle of CR to illuminate the entire vehicle. 
	\item[$\bullet$] \textbf{ISAC-based predictive beam tracking with alternant wide beam and narrow beam (ISAC-AB):}
	When the vehicle approaches the RSU, fewer antennas enlarging the beamwidth brings forth the declining array gain, resulting in decreased achievable rate. Motivated by this observation, a more advanced scheme is proposed by splitting each bolck into two parts. The first part is used for ISAC transmission, which has a dynamic beamwidth following ISAC-DB. The second part is dedicated to the communication performance, and transmitted on a narrow beam formulated by leveraging the estimated results from the first part. This results in a remarkable improvement in the achievable rate. Moreover, the inherent tradeoff between these two parts is revealed by a convex optimization model, where a time allocation strategy between the two stages is developed to optimize the average transmission rate.
\end{itemize}

The numerical results demonstrate that the proposed two schemes are capable of handling effective beam alignment, and providing more stable and reliable communication transmission over the conventional methods. It is worth highlighting that the more advanced ISAC-AB can attain excellent achievable rate in V2I networks.

The remainder of this article is organized as follows. Section II introduces the system modeling, Section III elaborates the proposed two schemes for beam trackinig, Section IV provides the simulations, and finally Section VI concludes the article.

\textit{\textbf{Notation:}} Throughout this paper, $\mathbf{A}$, $\mathbf{a}$, and $a$ denote matrix, vector, and scalar, respectively. 
$(\cdot)^T$, $(\cdot)^H$, $(\cdot)^{-1}$, $|\cdot|$, $\lfloor \cdot \rfloor$, $E(\cdot)$, $\mathbf{0}_{N}$, and $\mathbf{I}_{N}$ denote transport, Hermitian, inverse, modulus of a complex number, equal to or smaller than a real number, expectation, zero vector of size $N \times 1$, and identity matrix of size $N \times N$, respectively. $\arctan^{-1}(\cdot)$ denotes inverse tangent function in radian.
Besides, $\mathcal{N}(\bm{\mu},\mathbf{R})$ denotes Gaussian distribution with mean $\bm{\mu}$ and covariance matrix $\mathbf{R}$, while $\mathcal{CN}(\bm{\mu},\mathbf{R})$ represents circularly symmetric complex Gaussian distribution with mean $\bm{\mu}$ and covariance matrix $\mathbf{R}$. Finally, $\mathbb{C}^{N \times M}$ denotes the complex space of dimension $N \times M$.

\section{System Modeling} \label{sec2}

We consider a V2I downlink where a full-duplex RSU supports both sensing and data transmission. The RSU operates at the mmWave band and is equipped with mMIMO uniform linear array (ULA) which has $N_{t,n}$ transmit antennas and $N_r$ receive antennas\footnote{In practice, the RSU is installed with a known height, and the pitch angle is thus known \textit{a priori}. Hence, the considered 2D scenario with the ULA can be readily extended to the 3D case with the uniform plane array (UPA).}. The subscript of $N_{t,n}$ indicates that it may change at different epoches $n$ on account of the varying beamwidth, which is distinctly different from many previous works like \cite{liu2020TWC,yuan2020,liu2020tutorialpart1,liu2020tutorialpart2,liu2020tutorial}. As for the vehicle, it moves along a straight road, which is parallel to the ULA. The extension to a non-parallel case is straightforward by rotating the coordinate system. The vehicle is modeled as an extended target equipped with a single-antenna CR. Resolvable scatterers are uniformly distributed in the vehicle geometry. Hence, once the coordinates of these scatterers are localized, the coordinates of centroid can be uniquely determined. Accordingly, the position of CR would be inferred since its relative coordinates to the centroid can be known \textit{a priori}. 

Without loss of generality, denote the angle, the distance and the velocity of the vehicle's CR relative to the RSU's array by $\phi(t)$, $d(t)$ and $v(t)$. All these motion parameters are defined in the region $t \in [0,T]$, with $T$ being the maximum time duration of interest. In the tracking procedure, we discretize the time period $T$ into several small time-slots with a duration of $\Delta T$. Further, denote the motion parameters at the $n$th epoch by $\phi_n$, $d_n$ and $v_n$, where $\phi_n = \phi(n\Delta T)$, $d_n = d(n\Delta T)$ and $v_n = v(n\Delta T)$.
Note that the predictive beamforming is conducted once within each interval of $\Delta T$. To ensure there is no \textit{range migration} so that the well-known ``stop-go" radar model \cite{richards2014} can be used, $v_n\Delta T \leq \Delta r = \frac{c}{2B}$ must be satisfied, where $\Delta r$ and $B$ denote the range resolution and the bandwidth, respectively. For instance, mmWave signaling occupies a large bandwidth of $500$MHz and a high mobility vehicle moves in the speed of $20m/s$, then $\Delta T <0.015s$ is parameterized.

\subsection{Radar Signal Model} 
At the $n$th epoch, the RSU receives the echo contributed by the vehicle's $K$ resolved scatterers with $N_r$ receive antennas, expressed as
\begin{equation}\label{equ1}
	\begin{aligned}
		\mathbf{r}_n (t) =  \kappa_n \sqrt{p_n} \sum^{K}_{k=0}  & \beta_{k,n} e^{j2\pi\mu_{k,n} t}   \mathbf{b} \left(\theta_{k,n}\right) \mathbf{a}^H \left(\theta_{k,n}\right) \mathbf{f}_n  s_n \left(t-\tau_{k,n}\right) + \mathbf{z}_r(t),
	\end{aligned}
\end{equation}
where $\mathbf{r}_n (t) \in \mathbb{C}^{N_r \times 1}$; $\kappa_n = \sqrt{N_{t,n}N_r}$ is the array gain factor; $p_n$ denotes the transmitted power; $\mu_{k,n}$, $\beta_{k,n}$ and $\tau_{k,n}$ denote the Doppler frequency, the complex reflection coefficient and the round-trip delay of the $k$th scatterer at the $n$th epoch, respectively; $s_n(t)$ is the transmitted ISAC signal;
$\mathbf{z}_n(t) \in \mathbb{C}^{N_r \times 1}$ represents the complex additive white Gaussian noise with zero mean and variance of $\sigma^2$, i.e., $\mathbf{z}_r(t) \sim \mathcal{CN}(\mathbf{0}_{N_r},\sigma^2\mathbf{I}_{N_r\times 1})$. Herein, the transmit SNR is defined as $\frac{p_n}{\sigma^2}$.

Besides, $\mathbf{a}(\theta)$ and $\mathbf{b}(\theta)$ are transmit and receive steering
vectors of the RSU's ULA, in the forms of
	\begin{align}\label{equ2}
		\mathbf{a}(\theta) & = \frac{1}{\sqrt{N_{t,n}}} \left[1, e^{-j\pi \cos\theta}, \cdots, e^{-j\pi (N_{t,n}-1) \cos\theta} \right]^T, \\
		\mathbf{b}(\theta) & = \frac{1}{\sqrt{N_r}} \left[1, e^{-j\pi \cos\theta}, \cdots, e^{-j\pi (N_r-1) \cos\theta} \right]^T,
	\end{align}
where a standard half-wavelength antenna spacing for the ULA is assumed, and $\theta \in (0,\pi)$. Here, we emphasize that $\mathbf{a}(\theta)$ depends on $N_{k,n}$, which indicates that its size may change at different epoches. 
The beamforming vector $\mathbf{f}_n$ is designed by exploiting the predicted angle of the CR, given as
\begin{equation}
	\begin{aligned}
		\mathbf{f}_n = \mathbf{a} \left(\widehat{\phi}_{n|n-1}\right),
	\end{aligned}
\end{equation}
where $\mathbf{f}_n \in \mathbb{C}^{N_{t,n}\times 1}$ is dynamic since the size of $N_{t,n}$ is adjustable. Here, $\widehat{\phi}_{n|n-1}$ denotes the predicted angle of the CR at the $n$th epoch based on the $(n-1)$th measurement.
Throughout this paper, we aim to estimate and predict the angle $\phi_n$ of the CR\footnote{To be specific, in the proposed ISAC-DB, only the prediction of $\phi_n$ is used; while in the more advanced ISAC-AB, both the estimation and prediction of $\phi_n$ are used. See the next section for more details.}, so as to steer the beam accurately and guarantee a reliable communication performance.

\begin{figure}[!t]
	\centering
	\includegraphics[width=3.5in]{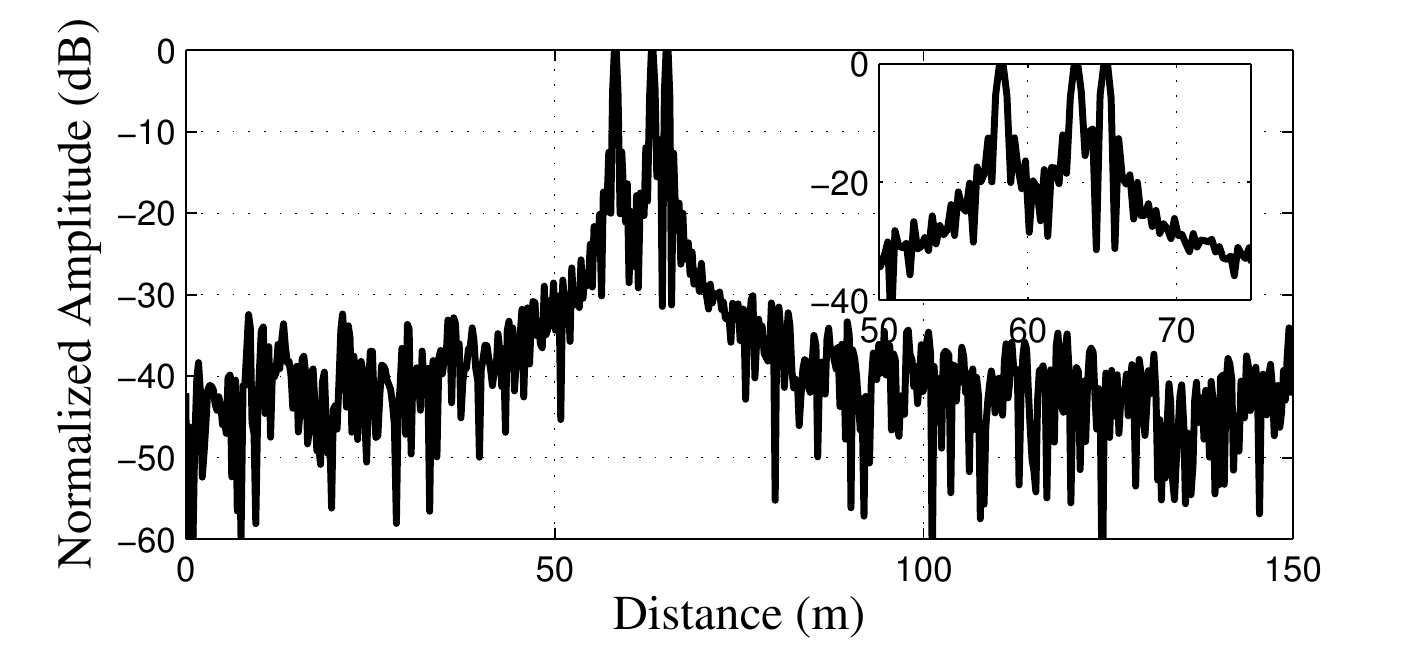}\\
	\caption{An example of matched-filtering output in the distance profiles. The signaling bandwidth is $500$MHz and the scenario contains three scatterers with distances as $(58,63,65)m$,  respectively. }\label{3.png}
\end{figure}

\subsection{Radar Measurement Model} First, we aim to determine the resolved ranges and velocities in the delay-Doppler domain by following a standard matched-filtering technique \cite{liu2020TWC}. Then the refined output is expressed as 
\begin{equation}\label{equ5}
	\begin{aligned}
		\widetilde{\mathbf{r}}_n = & \kappa_n \sqrt{p_n} \sum^{K}_{k=0}  \beta_{k,n} \mathbf{b} \left(\theta_{k,n}\right) \mathbf{a}^H \left(\theta_{k,n}\right)  \mathbf{a} \left(\widehat{\phi}_{n|n-1}\right)   \int^{\Delta T}_0 s_n \left(t-\tau_{k,n}\right) s^*_n \left(t-\tau \right) e^{-j2\pi (\mu-\mu_{k,n}) t}  dt   +\widetilde{\mathbf{z}}_r \\
		= & \kappa_n \sqrt{p_n} \sqrt{G} \sum^{K}_{k=0} \beta_{k,n} \mathbf{b} \left(\theta_{k,n}\right) \mathbf{a}^H \left(\theta_{k,n}\right)  \mathbf{a} \left(\widehat{\phi}_{n|n-1}\right)   \bar{\delta} \left(\tau-\tau_{k,n}; \mu-\mu_{k,n} \right)  +\widetilde{\mathbf{z}}_r,
	\end{aligned}
\end{equation}
where $\bar{\delta} (\tau;\mu)$ is the normalized matched-filtering output function obtained by time and frequency-reversing and conjugating its own waveform for the complex transmit signal $s_n(t)$ \cite{yi2020}. 
Notice that $\bar{\delta} (\tau;\mu)$ generally has a narrow-mainlobe property in both time domain and Doppler domain to ensure high distance and velocity resolution, and $\bar{\delta} (\tau;\mu)=1$ when $\tau=0$ and $\mu=0$.
The usage of matched-filtering enables the receive SNR of sensing to improve by a matched-filtering gain denoted as $G$.
Note that $G$ is equal to the number of symbols used for matched-filtering in a block. 
Besides, $\widetilde{\mathbf{z}}_r$ represents the noise output of matched-filtering. Hence, the distance and Doppler of the $k$th scatterer can be readily measured by finding the location of the corresponding peak position, expressed as 
	\begin{align}
		\widehat{\tau}_{k,n} = \frac{2d_{k,n}}{c} + z_{\tau_{k,n}}, \
		\widehat{\mu}_{k,n} = \frac{2v_n\cos(\theta_{k,n})f_c}{c} + z_{\mu_{k,n}}. 
	\end{align}

\textbf{\textit{Definition:}} In consideration of the radar signal with a large bandwidth, a narrow mainlobe of matched-filtering output function corresponds to a high range resolution. 
The $k$th and the $k'$th scatterers ($k, k'=1,2,\cdots,K$ and $k \neq k'$) are said to be separable in the time-delay axis, if the time difference of arrival between them satisfies
 \begin{equation}
	\begin{aligned}
		\left\vert \tau_{k,n}-\tau_{k',n} \right\vert > \tau_0, 
	\end{aligned}
\end{equation}
where $\tau_0$ is the effective mainlobe duration of the matched-filtering output function in time-delay axis (i.e., corresponding to the range resolution $\Delta r$). An example of matched-filtering output is shown in Fig. \ref{3.png}, where three scatterers are separable since their distance differences are larger than $\Delta r$.
In the Doppler domain there is a similar definition, which is omitted here for notational convenience.

For the $k$th scatterer, substituting $\tau=\widehat{\tau}_{k,n}$ and $\mu=\widehat{\mu}_{k,n}$ into the matched-filtering output function (\ref{equ5}) and relying on the above definition yields
\begin{equation}\label{cube}
	\begin{aligned}
		\widetilde{\mathbf{r}}_{k,n} = \kappa_n \beta_{k,n} \mathbf{b} \left(\theta_{k,n} \right) \mathbf{a}^H \left(\theta_{k,n} \right)  \mathbf{a} \left(\widehat{\phi}_{n|n-1} \right)  + \bar{\mathbf{z}}_r,
	\end{aligned}
\end{equation}
which indicates that $K$ scatterers are totally separated. Here, $\bar{\mathbf{z}}_r$ denotes the measurement noise normalized by the transmit power $p_n$ and the matched-filtering gain $G$, which follows $\bar{\mathbf{z}}_r \sim \mathcal{CN}(\mathbf{0}_{N_r},\sigma^2_\theta\mathbf{I}_{{N_r}\times 1})$, where $\sigma^2_\theta = \frac{\sigma^2}{Gp_n} $.
Besides, the complex reflection coefficient $\beta_{k,n}$ is determined by the signal propagation distance and the RCS of the vehicle, expressed as
\begin{equation}
	\begin{aligned}
\beta_{k,n} = \frac{\varepsilon_{k,n}}{(2d_{k,n})^2} = \frac{\varepsilon_{k,n}}{c^2\tau^{2}_{k,n}},
	\end{aligned}
\end{equation}
where $\varepsilon_{k,n}$ denotes the complex RCS of the $k$th slowly fluctuated Swerling I-type scatterer \cite{richards2014} at the $n$th epoch. 
The above thorough separation among scatterers is on the basis of the assumption that the sidelobes of $\bar{\delta} (t;\mu)$ are negligible. In practice, sidelobes would affect the perfect separation and their influence is not considered in this article.
Then $\theta_{k,n}$ can be readily measured by the maximum likelihood estimation (MLE) \cite{kayest} or super-resolution algorithms like multiple signal classification (MUSIC) \cite{keskin2021}, whose measurements are expressed as\footnote{In order to measure the angle, one must construct the sample covariance matrix in accordance with the space/time snapshot in (\ref{cube}), where at least $N_r$ samples are required in the $n$th block. See \cite{keskin2021} for more details.}
\begin{equation}
	\begin{aligned}
		\widehat{\theta}_{k,n} = \theta_{k,n} + z_{\theta_{k,n}}. 
	\end{aligned}
\end{equation}

Note that $z_{\theta_{k,n}}$, $z_{\tau_{k,n}}$ and $z_{\mu_{k,n}}$ are additive noises with zero means and variances of $\sigma^2_{k,n}(1)$, $\sigma^2_{k,n}(2)$ and $\sigma^2_{k,n}(3)$, respectively. Moreover, we remark here that the measurement variances are inversely proportional to the transmit SNR of (\ref{equ1}), i.e.,
\begin{equation}
	\sigma^2_{k,n}(i) = \frac{a^2_i\sigma^2}{p_n G|\kappa_n \beta_{k,n}|^2\left\vert \varrho_{k,n} \right\vert^2}, \ i=1,2,3.
\end{equation}
where $\varrho_{k,n} = \mathbf{a}^H(\theta_{k,n}) \mathbf{a} \left(\widehat{\phi}_{n|n-1} \right)$ represents the beamforming gain factor of $k$th scatterer, whose modulus is less than 1 since the beam points to the CR rather than the $k$th scatterer.
Finally, $a_i$, $i=1,2,3$ are constants related to the system configuration, signal designs as well as the specific signal processing algorithms \cite{liu2020TWC}.

Now, with the measured $\widehat{\tau}_{k,n}$ and $\widehat{\theta}_{k,n}$, the coordinates of vehicle centroid $(x_n,y_n)$ can be localized in the 2D Cartesian coordinate as
\begin{align}
	\widehat{x}_n = \frac{1}{K}\sum^{K}_{k=1}{c\widehat{\tau}_{k,n}\cos \widehat{\theta}_{k,n}}/2 = \frac{1}{K}\sum^{K}_{k=1}{\widehat{d}_{k,n}\cos \widehat{\theta}_{k,n}}, \
	\widehat{y}_n = \frac{1}{K}\sum^{K}_{k=1}{c\widehat{\tau}_{k,n}\sin \widehat{\theta}_{k,n}}/2 = \frac{1}{K}\sum^{K}_{k=1}{\widehat{d}_{k,n}\sin \widehat{\theta}_{k,n}}. 
\end{align}

Next, we estimate the velocity. Given that
\begin{equation}
\begin{aligned}
\widehat{\mu}_{k,n} = & \frac{2v_n\cos(\theta_{k,n})f_c}{c} + z_{\mu_{k,n}}, \ \forall k
\end{aligned}
\end{equation}
we reformulate $K$ equations with the matrix-vector form as
\begin{equation}
\begin{aligned}
\widehat{\bm{\mu}}_n = \mathbf{A}(\bm{\theta}_n) v_n + \mathbf{z}_{\mu},
\end{aligned}
\end{equation}
where $\widehat{\bm{\mu}}_n = \left[\widehat{\mu}_{1,n},\widehat{\mu}_{2,n},\cdots,\widehat{\mu}_{K,n} \right]^T$, $\mathbf{A}(\bm{\theta}_n) = \left[\frac{2f_c\cos(\theta_{1,n})}{c},\frac{2f_c\cos(\theta_{2,n})}{c},\cdots,\frac{2f_c\cos(\theta_{K,n})}{c} \right]^T$ and $\mathbf{z}_{\mu} = \left[z_{\mu_{1,n}},z_{\mu_{2,n}},\cdots,z_{\mu_{K,n}} \right]^T$.
It is straightforward to see that $\widehat{\bm{\mu}}_n\sim \mathcal{N}(\mathbf{A}(\bm{\theta}_n) v_n,\mathbf{Q}_{\mu})$, where $\mathbf{Q}_{\mu}=\text{diag}(\sigma^2_{1,n}(3),\cdots,\sigma^2_{K,n}(3))$. In view of this, the MLE of $v_n$ conditional to $\bm{\theta}_n$ is thus given as \cite{kayest}
\begin{equation} 
\begin{aligned}
\widehat{v}_{n|\bm{\theta}_n} & = \left(\mathbf{A}^T(\bm{\theta}_n) \mathbf{Q}^{-1}_{\mu}\mathbf{A}(\bm{\theta}_n) \right)^{-1}\mathbf{A}^T(\bm{\theta}_n) \mathbf{Q}^{-1}_{\mu}\widehat{\bm{\mu}}_n  = \frac{c}{2f_c} \cdot \frac{\sum^{K}_{k=1}\widehat{\mu}_{k,n}\cos(\theta_{k,n})/\sigma^2_{k,n}(3)}{\sum^{K}_{k=1}\cos^2(\theta_{k,n})/\sigma^2_{k,n}(3)}.
\end{aligned}
\end{equation}

\textbf{\textit{Assumption:}} Denote the coordinate of CR as $(x_n+\Delta x, y_n+\Delta y)$ relative to the RSU, where $\Delta x$ and $\Delta y$ are relative coordinates of the CR to the centroid. Note that $\Delta x$ and $\Delta y$ are known since they can be fed back to the RSU at the initial beam training when the vehicle is captured.

The measurement model is finally summarized as\footnote{Note that $\arctan^{-1}(x) \in (-\pi/2,\pi/2) $, while the angle of vehicle is defined in the region of $[0,\pi) $. For clarity, we define $\tan^{-1}\left( x \right)=\arctan \left( x \right)$ if $x\geq 0$, otherwise $\tan^{-1}\left( x \right)=\arctan \left( x \right) + \pi$.}
\begin{equation}\label{equ18}
	\left\{
	\begin{aligned}
	\widehat{\phi}_n & = \tan^{-1}\left( \frac{\widehat{y}_n + \Delta y} {\widehat{x}_n + \Delta x } \right) = \phi_n + z_\phi, \\
	\widehat{d}_n & = \sqrt{(\widehat{x}_n+\Delta x)^2 + (\widehat{y}_n+\Delta y)^2} = d_n + z_d, \\
	\widehat{v}_n & = \frac{c}{2f_c} \cdot \frac{\sum^{K}_{k=1}\widehat{\mu}_{k,n}\cos(\widehat{\theta}_{k,n})/\sigma^2_{k,n}(3)}{\sum^{K}_{k=1}\cos^2(\widehat{\theta}_{k,n})/\sigma^2_{k,n}(3)} = v_n + z_v,
	\end{aligned}
	\right.
\end{equation}
where $z_{\phi}$, $z_d$ and $z_v$ are measurement noises with zero means and variances of $\sigma^2_{\phi}$, $\sigma^2_d$ and $\sigma^2_v$, respectively. 
Most of the existing literature concerning the extended target tracking like \cite{salmond2003,godrich2010,Zhong2010spl} adopts the ideal assumption of known variances of measurement noises. In this article, while $\widehat{\theta}_{k,n}$, $\widehat{d}_{k,n}$ and $\widehat{v}_n$ are assumed Gaussian distributed, $\widehat{\phi}_n$, $\widehat{d}_n$ and $\widehat{v}_n$ are not Gaussian due to the non-linearity of (\ref{equ18}). More importantly, $\sigma^2_{\phi}$, $\sigma^2_d$ and $\sigma^2_v$ are unknown in practice. To proceed, we develop approaches for obtaining approximate variance counterparts. The detailed derivations are provided in the Appendix.

\subsection{Communication Receiver Model}
We emphasize that, the ISAC signal is used for both vehicle tracking and communication transmission during the entire transmission block. Hence, at the $n$th epoch, the vehicle receives the signal transmitted by the RSU as
\begin{equation}
	\begin{aligned}
		y^C_n(t) = \alpha_n \kappa^C_n \sqrt{p}  \mathbf{a}^H \left(\phi_n \right) \mathbf{a} \left(\widehat{\phi}_{n|n-1} \right) s_n(t) + z^C_n(t).
	\end{aligned}
\end{equation}
where $\kappa^C_n=\sqrt{N_{t,n}}$ is the array gain.
Besides, $\alpha_n$ indicates the LoS channel coefficient, which is given as \cite{liu2019} $\alpha_n = \alpha_\text{ref} d^{-1}_n e^{j\frac{2\pi f_c}{c}d_n}$, where $\alpha_\text{ref} d^{-1}_n$ is the LoS path-loss of the channel, and $\alpha_\text{ref}$ is a known reference path-loss measured at the distance of $d_0=1m$. Thus, estimating $\alpha_n$ is equivalent to estimating $d_n$.


The receive SNR of the CR is therefore given as $\text{SNR}^C_n = p_n \left\vert \kappa^C_n \alpha_n \right\vert^2 \left\vert \varrho^C_{k,n} \right\vert^2 /\sigma^2_C$, where $\varrho^C_n = \mathbf{a}^H(\phi_n) \mathbf{a}(\widehat{\phi}_{n|n-1}) $ represents the beamforming gain factor, whose modulus equals to 1 if the predicted
angle perfectly matches the real angle, and is less than
1 otherwise. Clearly, the angle prediction accuracy of CR is predominating in $\text{SNR}^C_n$.
Finally, the achievable rate at the $n$th epoch is formulated as
\begin{equation}\label{equ22}
	\begin{aligned}
		R_n = \log \left(1 + \text{SNR}^C_n \right).
	\end{aligned}
\end{equation}

\textbf{\textit{Remark:}} The synchronization of the reveived communication signal can be achieved with pilots at the beginning of the frame structure. Besides, the achievable rate (\ref{equ22}) refers to the upper bound of the practical communication rate.


\section{ISAC-based Predictive Beam Tracking Schemes for V2I Linking}
As discussed previously, our main focus is to track the CR. For this purpose, the key premise is that the beam must be able to cover the entire vehicle in physical size, in order to cover resolved scatterers as many as possible and to accurately refine the position of centroid. Therefore, the beamwidth must be adjusted in real-time according to the vehicle's trajectory.

We refer readers to \cite{van2004} for a detailed definition of the ULA beamwidth.
Here, instead of the first null beamwidth, we resort to the commonly-used half-power beamwidth for the ULA, given as 
\begin{equation}\label{equ23}
	\begin{aligned}
		\theta_{BW} =  \frac{k_0c}{N_td_0f_c} \cdot \frac{1}{\sin(\phi)} \approx  \frac{1.78}{N_{t,n}\sin(\phi)},
	\end{aligned} 
\end{equation}
where $k_0$, $d_0$, $f_c$ and $\phi$ denote the beamwidth factor, the ULA spacing, the carrier frequency and the angle of the CR, respectively. Here, the approximation holds when a half wavelength spacing is employed for the ULA.  

The coverage width $\Delta d$ can thus be approximately calculated by a trigonometric function as 
\begin{equation}\label{equ24}
	\begin{aligned}
		\Delta d  = 2d_n\cdot \tan\left( \frac{\theta_{BW}}{2} \right) \approx 2d\cdot \tan\left( \frac{0.89}{N_{t,n}\sin(\phi)} \right).
	\end{aligned} 
\end{equation}
In our considered scenario, the vehicular coverage width $\Delta d$ is assumed to be constant. For example, in terms of a vehicle of the length $5$m and the width $2m$, the maximum coverage width can be set as $\Delta d > \sqrt{5^2+2^2}=5.385m$.

Moreover, since the beamwidth varies at different epoches, the number of transmit antennas is calculated in terms of (\ref{equ24}) based on the predicted $\widehat{d}_{n|n-1}$ and $\widehat{\phi}_{n|n-1}$.
Besides, we emphasize that the mMIMO system allows a maximum antenna number as $N_{t,\text{max}}$ (e.g., $N_{t,\text{max}}=128$). As a consequence, the number of transmit antennas is summarized as
\begin{equation}\label{equ25}
	\begin{aligned}
		N_{t,n} = \min \left\{ \left \lfloor \frac{0.89}{ \arctan^{-1} \left( \frac{\Delta d}{2\widehat{d}_{n|n-1}} \right) \cdot \sin(\widehat{\phi}_{n|n-1}) } \right \rfloor, N_{t,\text{max}} \right\}.
	\end{aligned} 
\end{equation}


\subsection{ISAC-DB Method}
\begin{figure}[!t]
	\begin{tabular}{cc}
		\begin{minipage}[t]{0.48\linewidth}
			\includegraphics[width=3.225in]{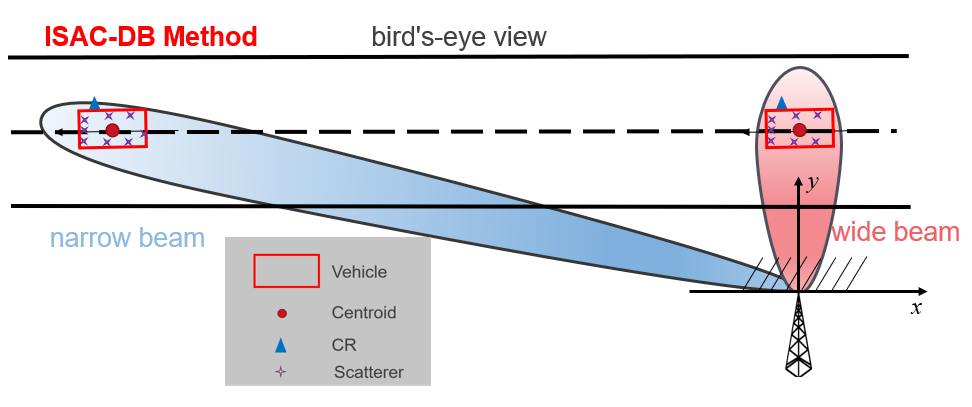}\\
			\caption{Scenario of ISAC-DB: a benchmark design with the dynamic beamwidth for beam tracking.}\label{figg3}
		\end{minipage}
		\begin{minipage}[t]{0.48\linewidth}
			\includegraphics[width=3.35in]{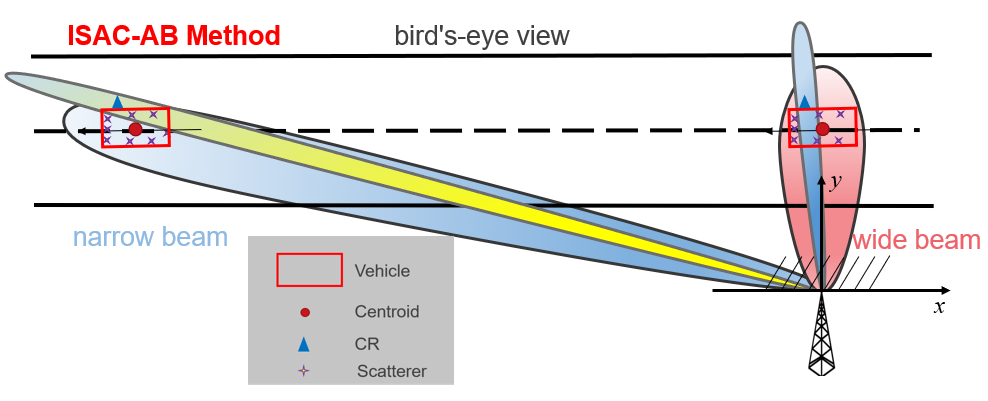}\\
			\caption{Scenario of ISAC-AB: a more advanced design with alternant wide beam and narrow beam for beam tracking.}\label{figg4}
		\end{minipage}
	\end{tabular}
\end{figure}

A scenario diagram is illustrated in Fig. \ref{figg3}.
At the $n$th epoch, the state vector is composed of the angle, the distance and the velocity of the CR, given as
\begin{equation}
	\begin{aligned}
		\mathbf{x}_n = (\phi_n,d_n,v_n)^T.
	\end{aligned}
\end{equation}

Following the derivation in \cite{liu2020TWC}, it is straightforward to summarize the state evolution model of the CR as
\begin{equation}
	\left\{
	\begin{aligned} 
		\phi_n & = \phi_{n-1} + d^{-1}_{n-1}v_{n-1}\Delta T \sin(\phi_{n-1}) + \omega_\phi, \\
		d_n & = d_{n-1} - v_{n-1}\Delta T \cos(\phi_{n-1}) + \omega_d, \\
		v_n & = v_{n-1} + \omega_v.
	\end{aligned} 
	\right. 
\end{equation}

The state evolution model and the measurement model are reformulated by more compact matrix-vector forms as
\begin{equation}
	\left\{
	\begin{aligned} 
		\text{State Evolution Model:} \ \mathbf{x}_n & = \mathbf{h} (\mathbf{x}_{n-1}) +  \bm{\omega}_n , \\
		\text{Measurement Model:} \ \mathbf{y}_n & = \mathbf{x}_n +  \mathbf{z}_n ,
	\end{aligned} 
	\right. 
\end{equation}
where $\mathbf{y}_n = \widehat{\mathbf{x}}_n = [\widehat{\phi}_n, \widehat{d}_n,\widehat{v}_n]^T$, 
$\bm{\omega}_n=[\omega_\phi,\omega_d,\omega_v]^T$ and $\mathbf{z}_n = [z_\phi,z_d,z_v]^T$. 
As considered above, both $\bm{\omega}_n$ and $\mathbf{z}_n$ are of covariance matrices being expressed as
\begin{align}
	\mathbf{Q}_w = \text{diag} \left( \bar{\sigma}^2_\phi,\bar{\sigma}^2_d,\bar{\sigma}^2_v \right), \
	\mathbf{Q}_z = \text{diag} \left( \sigma^2_\phi,\sigma^2_d,\sigma^2_v \right).
\end{align} 


Because of the non-linearity of the state evolution model, the standard Kalman filtering can not be directly utilized. Therefore, EKF is used by linearizing the state evolution model, where the Jacobian matrix should be derived as
\begin{equation}
	\begin{aligned}
		\mathbf{H} = \frac{\partial\mathbf{h}} {\partial \mathbf{x}} = 
		\left[ \begin{array}{cccc}
			1+ \frac{v \Delta T \cos(\phi)}{d} & -\frac{v \Delta T \sin(\phi)}{d^2} & 0 \\
			v\Delta T \sin(\phi) & 1 & -\Delta T \cos(\phi) \\
			0 & 0 & 1
		\end{array}
		\right] .
	\end{aligned} 
\end{equation}

Before presenting the EKF technique, the remaining problem is the initialization of algorithm. In practice, it can be achieved with a conventional beam training with limited pilots at the beginning of the frame structure, then $\widehat{\phi}_0$, $\widehat{d}_0$ and $\widehat{v}_0$ can be generated. Following the standard procedure of EKF \cite{kayest} with \textbf{\textit{Step 1)-6)}}, the state prediction and tracking are summarized in Algorithm 1. 
\begin{algorithm}[!t]
	\caption{Predictive beam tracking of ISAC-DB.}
	\begin{algorithmic}[1] 
		\STATE  {\textbf{\textit{Step 1)}} State prediction:} $\widehat{\mathbf{x}}_{n|n-1}  = \mathbf{h} \left(\widehat{\mathbf{x}}_{n-1} \right)$.
		\STATE  {\textbf{\textit{Step 2)}} Linearization:} $
			\mathbf{H}_{n-1} = \frac{\partial\mathbf{h}}{\partial \mathbf{x}} \bigg|_{\mathbf{x}=\widehat{\mathbf{x}}_{n-1}}$.
		\STATE  {\textbf{\textit{Step 3)}} Mean square error (MSE) matrix prediction:} $\mathbf{M}_{n|n-1} = \mathbf{H}_{n-1}\mathbf{M}_{n-1}\mathbf{H}^H_{n-1} + \mathbf{Q}_{\omega}$.
		\STATE  {\textbf{\textit{Step 4)}} Kalman gain calculation:} $\mathbf{K}_{n} = \mathbf{M}_{n|n-1}  \left(\mathbf{M}_{n|n-1}+\mathbf{Q}_z \right)^{-1}$.
		\STATE  {Transmit antenna number update with (\ref{equ25}).}
		\STATE  {Measuremet update: $\mathbf{y}_{n} = \mathbf{x}_{n} + \mathbf{z}_{n}$.}
		\STATE  {\textbf{\textit{Step 5)}} State tracking:} $\widehat{\mathbf{x}}_{n} = \widehat{\mathbf{x}}_{n|n-1} + \mathbf{K}_{n} \left( \mathbf{y}_{n} - \widehat{\mathbf{x}}_{n|n-1} \right)$.
		\STATE  {\textbf{\textit{Step 6)}} MSE matrix update:} $\mathbf{M}_{n} = \left( \mathbf{I}_3-\mathbf{K}_{n} \right) \mathbf{M}_{n|n-1}$.
	\end{algorithmic}
\end{algorithm}

%
%
%
%
%
%

\begin{figure}[!t]
	\centering
	\includegraphics[width=3.5in]{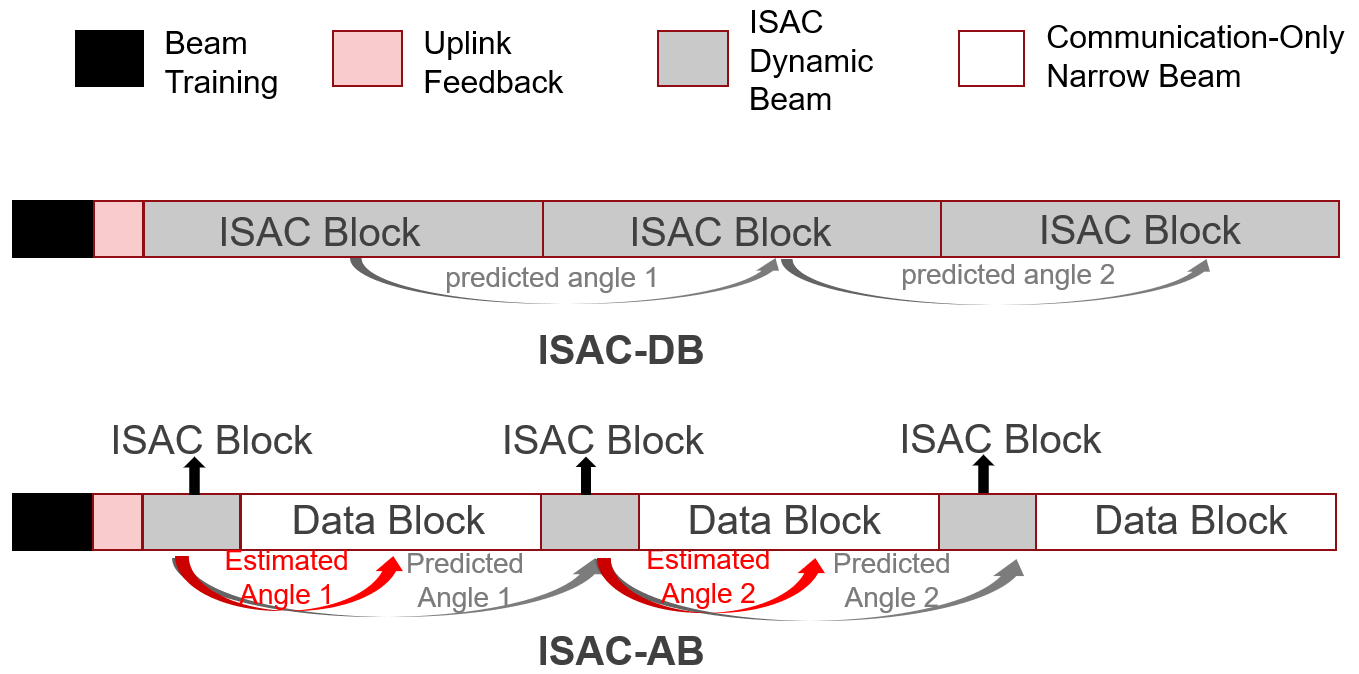}\\
	\caption{Frame structures of the proposed schemes.}\label{fig5}
\end{figure}

\subsection{ISAC-AB Method}

ISAC-DB requires the entire coverage of the vehicle with a dynamic wide beam, so that the position of the CR can be inferred relying on the resolved extended target profile. However, when the vehicle approaches the RSU, the wider beam adopted deteriorates the communication performance due to the lower array gain. To balance such a contradiction, we further design a more advanced scheme wherein each block is divided into two parts: the former part transmits a wide beam for ISAC transmission, while the latter part is solely for communication by using a narrow beam which aligns the CR with the angle estimated from the former part. This is because the narrow beam is incapable of covering the entire vehicle geometry, thus the position of the CR can not be inferred according to the resolved geometry information. A scheme diagram is illustrated in Fig. \ref{figg4}. 
Besides, we also refer readers to Fig. \ref{fig5} for more distinct expressions of the frame structures.

Denote the achievable rates of the wide beam and the narrow beam by $R^\text{wide}_n$ and $R^\text{narrow}_n$, respectively. We introduce a time splitting factor as $\rho$, then the goal we wish to achieve is to maximize the average achievable rate in each block. Since only the former part in each block is used for sensing, $\rho$ has a similar interpretation to the concept of duty ratio in each pulse repetition interval (PRI) \cite{richards2014}. To elaborate, in the $n$th epoch, the optimization model is formulated as 
\begin{equation}\label{equ38}
	\begin{aligned}
		& \rho_\text{opt} = \arg \max_{\rho} \ \rho R^\text{wide}_n + (1-\rho)R^\text{narrow}_n \\
		& s.t. \ 0<\rho \leq 1.
	\end{aligned}
\end{equation}

\textit{Interpretation of the tradeoff relationship:}
When beams are aligned, it is apparent that $R^\text{narrow}_n \gg R^\text{wide}_n$ due to the much higher array gain of narrow beam, so that one aims to make $\rho$ as small as possible, i.e., to allocate longer time duration to the narrow beam. However, an excessively small $\rho$ may result in a failed tracking, which yields beam misalignment. In this case, the second term of the objective function becomes zero\footnote{In fact, beam sidelobes towards the CR still have an impact on the communication performance. However, the resulting achiveable rate is small enough and can be approximately ignored.}. However, the CR can not be outside the wide beam such that $R^\text{wide}_n>0$ always holds.
Inherently, there is a noteworthy tradeoff relationship in such an interesting problem. The remaining task is to find the optimum solution of $\rho$.

Now, denote the beam alignment probability for the narrow beam by $P_\text{A}$. Then the average achievable rate is formulated as
\begin{equation}\label{equ39}
	\begin{aligned}
		R^\text{narrow}_n & = \int_{\Omega} R p(R ) dR  \approx P_\text{A} R^\text{narrow}_{\text{align},n} + (1-P_\text{A}) R^\text{narrow}_{\text{mis},n},
	\end{aligned}
\end{equation}
where $ R \in \Omega \triangleq  R^\text{narrow}_{\text{align},n} \cup R^\text{narrow}_{\text{mis},n} $, and $R^\text{narrow}_{\text{align},n}$ and $R^\text{narrow}_{\text{mis},n}$ represent the achievable rates of beam alignment and beam misalignment for the narrow beam, respectively. 

To summarize, $R^\text{wide}_n$, $R^\text{narrow}_{\text{align},n}$ and $R^\text{narrow}_{\text{mis},n}$ are expressed as
	\begin{align}				
		R^\text{wide}_n & = \log\left( 1+p_n\frac{\left\vert \alpha_n\kappa^C_n \mathbf{a}^H(\phi_n)\mathbf{a}(\widehat{\phi}_{n|n-1}) \right\vert^2}{\sigma^2_c} \right),\label{equ40} \\
		R^\text{narrow}_{\text{align},n} & = \log\left( 1+p_n\frac{\left\vert \alpha_n \bar{\kappa}^C \bar{\mathbf{a}}^H(\phi_n) \bar{\mathbf{a}}(\widehat{\phi}_n) \right\vert^2}{\sigma^2_c} \right), \label{equ41}\\
		R^\text{narrow}_{\text{mis},n} & = 0, \label{equ42}
\end{align}
where the beamforming vector $\mathbf{f}^\text{narrow}_n$ for the narrow beam is used in (\ref{equ41}), which is designed by exploiting the estimated results, given as
\begin{equation}
	\begin{aligned}
		\mathbf{f}^\text{narrow}_n = \bar{\mathbf{a}} \left(\widehat{\phi}_n \right).
	\end{aligned}
\end{equation}
Note that $\bar{\kappa}^C$ and $\bar{\mathbf{a}}(\phi)$ are totally different from $\kappa^C_n$ and $\mathbf{a}(\phi)$. For $\bar{\kappa}^C$ and $\bar{\mathbf{a}}(\phi)$, they are fixed with a constant number of mMIMO antennas. Here, we have 
\begin{equation}
	\begin{aligned}
		\bar{\mathbf{a}}(\phi)=\frac{1}{\sqrt{N^\text{narrow}_t}} \left[1, e^{-j\pi \cos\phi}, \cdots, e^{-j\pi (N^\text{narrow}_t-1) \cos\phi} \right]^T, 
	\end{aligned}
\end{equation}
and 
\begin{equation}
	\begin{aligned}
		\bar{\kappa}^C=\sqrt{N^\text{narrow}_t}, 
	\end{aligned}
\end{equation}
where $N^\text{narrow}_t$ denotes the antenna number of the narrow beam and is constant according to the system configuration. Thus, $\bar{\kappa}^C$ has no subscript. As for $\kappa^C_n$ and $\mathbf{a}(\phi)$, they are dynamic because the beamwidth of the wide beam varies to cover the entire vehicle geometry.
Both the predicted angle from the last epoch (i.e., $\widehat{\phi}_{n|n-1}$) and the estimated angle from the current epoch (i.e., $\widehat{\phi}_n$) are combined in the objective function. It is highlighted that $\widehat{\phi}_{n|n-1}$ is irrelevant to the current $\rho$ at the $n$th epoch, while $\widehat{\phi}_n$ is relevant. 

Substiting (\ref{equ39}), (\ref{equ40}), (\ref{equ41}) and (\ref{equ42}) into (\ref{equ38}) yields 
\begin{equation}
	\begin{aligned}
		& \rho_\text{opt} = \arg \max_{\rho} \ \rho \log\left( 1+p_n\frac{\left\vert \alpha_n\kappa^C_n \mathbf{a}^H(\phi_n)\mathbf{a}(\widehat{\phi}_{n|n-1}) \right\vert^2}{\sigma^2_c} \right)  + (1-\rho) P_\text{A} \log\left( 1+p_n\frac{\left\vert \alpha_n \bar{\kappa}^C \bar{\mathbf{a}}^H(\phi_n) \bar{\mathbf{a}}(\widehat{\phi}_n) \right\vert^2}{\sigma^2_c} \right) \\
		& s.t.\ 0<\rho \leq 1.
	\end{aligned}
\end{equation}

To proceed, it is assumed that $\widehat{\phi}_n$ follows a Gaussian distribution around the true angle $\phi_n$ as $\widehat{\phi}_n \sim \mathcal{N}(\phi_n,\lambda^2_n)$. In ISAC-AB, the matched-filtering gain of the wide beam in the first part decreases by $\rho G$ times since the matched-filtering gain equals to the number of symbols in a block \cite{liu2020TWC}. If we let $\sigma^2_{\phi}$ represent the angle variance when $\rho=1$ as discussed before in Sec. III-A, then $\lambda^2_n=\sigma^2_{\phi}/\rho$.
The narrow beam is aligned only if the estimated angle in the wide beam satisfies $\left|\widehat{\phi}_n-\phi_n \right|<\delta_n$, where $\delta_n$ denotes the half-beamwidth of narrow beam defined in (\ref{equ23}), i.e.,
\begin{equation}
\begin{aligned}
\delta_n = \frac{0.89}{N^\text{narrow}_t \sin(\widehat{\phi}_n)} .
\end{aligned}
\end{equation}
However, the dependence of $\delta_n$ on $\widehat{\phi}_n$ is not benefical to the following derivations. Specifically, the estimation accuracy of $\widehat{\phi}_n$ is affected by the current $\rho_\text{opt}$ while we hope to obtain $\rho_\text{opt}$ according to the known $\widehat{\phi}_n$. To proceed, we use the predicted $\widehat{\phi}_{n|n-1}$ instead because its prediction accuracy is determined by the last $\rho$ and is irrelevant to the current $\rho$. In brief, in the optimization model we now use 
\begin{equation}
\begin{aligned}
\delta_n \approx  \frac{0.89}{N^\text{narrow}_t \sin(\widehat{\phi}_{n|n-1})}.
\end{aligned}
\end{equation}
Likewise, despite the true value of $\alpha_n$ being $\alpha_n = \alpha_\text{ref} d^{-1}_n e^{j\frac{2\pi f_c}{c}d_n}$ which is unknown in practice, we use $\widehat{\alpha}_{n|n-1} = \alpha_\text{ref} \widehat{d}^{-1}_{n|n-1} e^{j\frac{2\pi f_c}{c}\widehat{d}_{n|n-1}}$ in the optimization for the same reason, for both the wide beam and the narrow beam.

Now, the task left is to derive the beam alignment probability of the narrow beam, which is given as
\begin{equation}
	\begin{aligned}
		P_\text{A} = & \text{Pr} \left\{ \widehat{\phi}_n\leq \phi_n + \delta_n \right\} - \text{Pr} \left\{ \widehat{\phi}_n\leq \phi_n - \delta_n \right\} \\
		= & \frac{1}{2} \left( 1 + \text{erf}\left( \frac{\delta_n}{\sqrt{2}\lambda_n} \right)\right) - \frac{1}{2} \left( 1 + \text{erf}\left( \frac{-\delta_n}{\sqrt{2}\lambda_n} \right)\right) \\
		= & \text{erf}\left( \frac{\delta_n}{\sqrt{2}\lambda_n} \right),
	\end{aligned}
\end{equation}
where $\text{erf}(\cdot)$ represents the error function and $\text{Pr}(X\leq x)=\int^x_{-\infty}p_X(y)dy$.
Accordingly, the optimization problem is further recast as
\begin{equation}\label{equ47}
	\begin{aligned}
		 & \rho_\text{opt}= \arg \max_{\rho} \ \rho \log\left( 1+p_n\frac{\left\vert \widehat{\alpha}_{n|n-1}\kappa^C_n \mathbf{a}^H(\phi_n)\mathbf{a}(\widehat{\phi}_{n|n-1}) \right\vert^2}{\sigma^2_c} \right) \\ & \qquad + (1-\rho)\text{erf}\left( \sqrt{\frac{\rho}{2}}\frac{\delta_n}{\sigma_{\phi}} \right) \log\left( 1+p_n\frac{\left\vert \widehat{\alpha}_{n|n-1} \bar{\kappa}^C \bar{\mathbf{a}}^H(\phi_n) \bar{\mathbf{a}}(\widehat{\phi}_n) \right\vert^2}{\sigma^2_c} \right) \\
		& s.t.\ 0<\rho \leq 1.
	\end{aligned}
\end{equation}
Unfortunately, it is still impossible to obtain 
$\phi_n$ and $\widehat{\phi}_n$ prior to optimizating procedure at the $n$th epoch. For a practical application, 
$\left\vert \mathbf{a}^H(\phi_n)\mathbf{a}(\widehat{\phi}_{n|n-1}) \right\vert= \left\vert \frac{\sin\left(\pi N_{t,n}\left(\cos\phi_n-\cos\widehat{\phi}_{n|n-1}\right)\right)}{N_{t,n}\sin\left(\pi\left(\cos\phi_n-\cos\widehat{\phi}_{n|n-1}\right)\right)} \right\vert=1$
and $\left\vert \bar{\mathbf{a}}^H(\phi_n)\bar{\mathbf{a}}(\widehat{\phi}_n) \right\vert=  \left\vert \frac{\sin\left(\pi N^\text{narrow}_t\left(\cos\phi_n-\cos\widehat{\phi}_n\right)\right)}{N^\text{narrow}_t\sin\left(\pi\left(\cos\phi_n-\cos\widehat{\phi}_n\right)\right)} \right\vert= 1$ are assumed since beam alignments are reached, leading to a \textit{suboptimum} solution. 
Finally, the considered model is simplified as
\begin{equation}\label{equ48}
\begin{aligned}
& \rho_\text{opt} = \arg \max_{\rho} \ \rho \log\left( 1+p_n\frac{\left\vert \widehat{\alpha}_{n|n-1} \kappa^C_n \right\vert^2}{\sigma^2_c} \right)  + (1-\rho)\text{erf}\left( \sqrt{\frac{\rho}{2}}\frac{\delta_n}{\sigma_{\phi}} \right) \log\left( 1+p_n\frac{\left\vert \widehat{\alpha}_{n|n-1} \bar{\kappa}^C \right\vert^2}{\sigma^2_c} \right) \\
& s.t.\ 0<\rho \leq 1.
\end{aligned}
\end{equation}

\textbf{\textit{Theorem:}} The model (\ref{equ48}) is a convex optimization problem.

\begin{figure}[!t]
	\centering
	\includegraphics[width=3in]{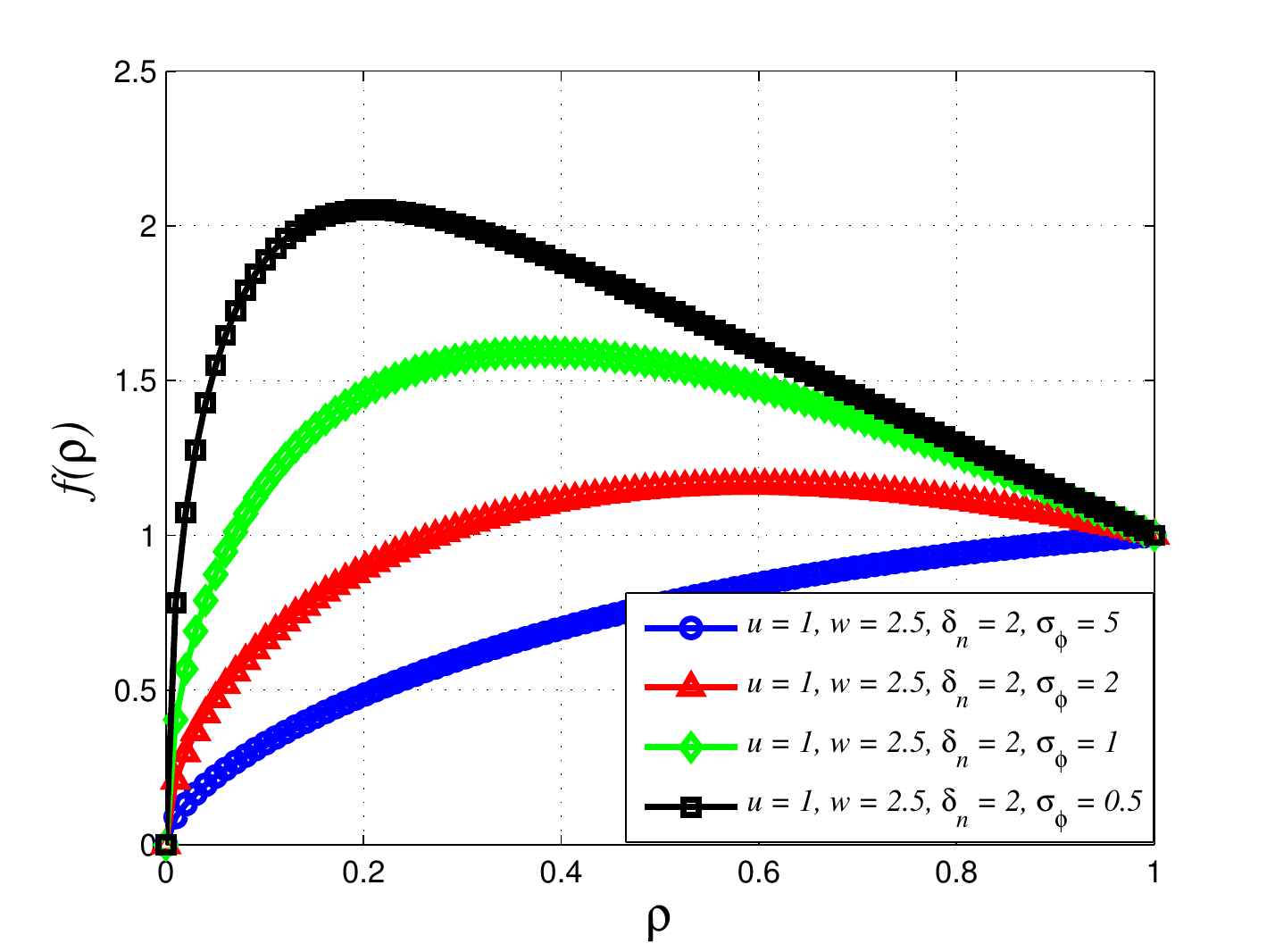}\\
	\caption{The objective function in (\ref{equ48}) $f(\rho)$ versus $\rho$.}\label{fig4}
\end{figure}

\textbf{\textit{Proof:}} To elaborate, we denote $u \triangleq \log\left( 1+p_n\frac{\left\vert \alpha_n\kappa^C_n \right\vert^2}{\sigma^2_c} \right) >0$, $v \triangleq \frac{\delta_n}{\sqrt{2}\sigma_{\phi}}>0$, and $w \triangleq \log\left( 1+p_n\frac{\left\vert \alpha_n \bar{\kappa}^C \right\vert^2}{\sigma^2_c} \right) >0$. Here, $u$, $v$ and $w$ are independent of $\rho$. 
Denote the objective function in (\ref{equ48}) by $f(\rho)$. Then the first-order derivative with respect to $\rho$  is derived as 
\begin{equation}\label{derivative1}
	\begin{aligned}
		f'(\rho) = u + \frac{wv}{\sqrt{\pi}} (\rho^{-\frac{1}{2}}-\rho^{\frac{1}{2}})e^{-\rho v^2} - w\text{erf}(\sqrt{\rho}v).
	\end{aligned}
\end{equation}
Therefore, the second-order derivative is further derived as
\begin{equation}
	\begin{aligned}
		f''(\rho) =  -\frac{wv}{\sqrt{\pi}}e^{-v^2\rho} \left( \frac{3}{2}\rho^{-\frac{1}{2}} + \frac{1}{2}\rho^{-\frac{3}{2}} + v^2\rho^{-\frac{1}{2}} - v^2\rho^{\frac{1}{2}} \right) < 0,
	\end{aligned}
\end{equation}
where $\partial \text{erf}(x)/\partial x = 2/\sqrt{\pi} e^{-x^2}$ and $0<\rho \leq 1$ are used. 
The proof is thus completed in terms of the negative second-order derivative and the linear feasible set. To vividly illustrate the convexity of (\ref{equ48}), numerical curves of $f(\rho)$ for $0<\rho \leq 1$ are provided in Fig. \ref{fig4}. Obviously, all curves are concave while the optimization is to find the maximum value. Hence, the convexity of (\ref{equ48}) is verified. Additionally, $\rho_\text{opt}$ is enlarged with the increasing $\sigma^2_{\phi}$, which indicates that more time durations for the wide beam is needed to ensure the beam alignment for the narrow beam in the low-SNR regime.

To solve problem (\ref{equ48}), let us define its Lagrangian multiplier as
\begin{equation}
	\begin{aligned}
		\mathcal{L}(\rho,\eta,\varsigma) = -f(\rho) - \eta\rho + \varsigma(\rho-1),
	\end{aligned}
\end{equation}
where $\eta \geq 0$ and $\varsigma \geq 0$.
According to \cite{boyd2004}, the Karush-Kuhn-Tucker (KKT) condition of (\ref{equ48}) is derived as
\begin{equation}
	\left\{
	\begin{aligned}
		& \mathcal{L}'(\rho_\text{opt},\eta^*,\varsigma^*) = -f'(\rho_\text{opt}) - \eta^*+\varsigma^* = 0, \\
		& \eta^* \geq 0, \ \varsigma^* \geq 0, \  
		\eta^* \rho_\text{opt} = 0, \ 
		\varsigma^* (\rho_\text{opt}-1)=0, \ 0<\rho_\text{opt}\leq 1.
	\end{aligned}
    \right.
\end{equation}
It is readily to see that the optimum solution is calculated as
$
		\rho_\text{opt} =\left\{
		\begin{array}{rcl}
			1      &     & {f'(1) \geq 0}\\
			\rho^*    &     & {\text{otherwise}}
		\end{array} \right. 
$, where $\rho^*$ is the sole root of $f'(\rho^*)=0$, which can be achieved by a trivial grid search.

Finally, the detailed steps of ISAC-AB can be summarized in Algorithm 2.
\begin{algorithm}[!http]
	\caption{Predictive beam tracking of ISAC-AB.}
	\begin{algorithmic}[1] 
		\FOR {$n=1,2,\cdots$}
		\STATE  {At the $n$th epoch, solve (\ref{equ48}) to obtain $\rho_\text{opt}$.}
		\STATE  {Beam tracking with the wide beam in $t\in [n\Delta T,(n+\rho_\text{opt})\Delta T)$: Following the steps in Algorithm 1, where the beamforming vector is designed with the predicted angle from the last epoch as} $\mathbf{f}_n = \mathbf{a} \left(\widehat{\phi}_{n|n-1}\right).$
		\STATE  {Communicating with the narrow beam in $t\in [(n+\rho_\text{opt})\Delta T,(n+1)\Delta T)$: The beamforming vector is designed with the estimated angle from the current epoch as} $\bar{\mathbf{f}}_n = \bar{\mathbf{a}} \left(\widehat{\phi}_n \right).$
	    \ENDFOR
	\end{algorithmic}
\end{algorithm}

We highlight that (\ref{equ48}) is a relaxation of the original problem and therefore its solution is suboptimal. By solving (\ref{equ48}), $\rho_\text{opt}$ is obtained and thus the real achievable rate at the $n$th epoch is given as
\begin{equation}\label{equ50}
	\begin{aligned}
		R^\text{opt}_n = \rho_\text{opt} \log\left( 1+p_n\frac{\left\vert \alpha_n\kappa^C_n \mathbf{a}^H(\phi_n)\mathbf{a}(\widehat{\phi}_{n|n-1}) \right\vert^2}{\sigma^2_c} \right) + (1-\rho_\text{opt}) \log\left( 1+p_n\frac{\left\vert \alpha_n \bar{\kappa}^C \bar{\mathbf{a}}^H(\phi_n) \bar{\mathbf{a}}(\widehat{\phi}_n) \right\vert^2}{\sigma^2_c} \right).
	\end{aligned}
\end{equation}

For comparison in subsequent simulations, the value of optimal objective function of (\ref{equ48}) at the $n$th epoch is also given as
\begin{equation}\label{equ51}
	\begin{aligned}
		R^\text{obj,opt}_n = \rho_\text{opt} \log\left( 1+p_n\frac{\left\vert \widehat{\alpha}_{n|n-1} \kappa^C_n \right\vert^2}{\sigma^2_c} \right)  + (1-\rho_\text{opt}) \text{erf}\left( \sqrt{\frac{\rho_\text{opt}}{2}}\frac{\delta_n}{\sigma_{\phi}} \right) \log\left( 1+p_n\frac{\left\vert \widehat{\alpha}_{n|n-1} \bar{\kappa}^C \right\vert^2}{\sigma^2_c} \right).
	\end{aligned}
\end{equation}
Note that $R^\text{obj,opt}_n $ is actually the approximately optimal average achievable rate since the effects of $|\mathbf{a}^H(\phi_n)\mathbf{a}(\widehat{\phi}_{n|n-1})|^2$ and $|\bar{\mathbf{a}}^H(\phi_n) \bar{\mathbf{a}}(\widehat{\phi}_n)|^2$ are ingored in (\ref{equ51}). Therefore, there would be a rate gap between $R^\text{obj,opt}_n$ and $R^\text{opt}_n$, especially when measurement variances are relatively large. A special case of $R^\text{obj,opt}_n \approx R^\text{opt}_n$ holds when measurement variances are small enough, in which both the wide beam and the narrow beam are accurately aligned, i.e., $|\mathbf{a}^H(\phi_n)\mathbf{a}(\widehat{\phi}_{n|n-1})|^2 \approx 1$, $|\bar{\mathbf{a}}^H(\phi_n) \bar{\mathbf{a}}(\widehat{\phi}_n)|^2 \approx 1$ and $P_\text{A} \approx 1$.


%

\begin{table}[!t]
	\centering
	\fontsize{10.0}{3.0}\selectfont
	\caption{Parameters in Simulations.}\label{tab2}
	\renewcommand\arraystretch{2}
	\begin{threeparttable}
		{\begin{tabularx}{7.0cm}{llll}
				\hline 
				\hline 
				Parameter & Value & Parameter & Value \\
				\hline
				$T$ & $8.0s$ &
				$\Delta T$ & $0.01s$ \\
				$f_c$ & $30$GHz &				
				$p_n$	& $1$ \\	
				$\Delta d$	& $6m$ &	
				$v$	& $20m/s$ \\
				$\Delta x$ & $1.5m$ &	
				$\Delta y$ & $0.5m$	\\
				$N^\text{narrow}_t$ & $128$ & $N_{t,\text{max}}$ & $128$ \\
				$a_1$ & $1.05e^{-2}$ &
				$a_2$ & $3.5e^{-2}$	\\
				$a_3$ & $1.05e^{-2}$ &	
				$N_r$ & $128$ \\
				$K$ & $8$ &					
				$\bar{\sigma}_\phi$ & $0.01^\circ$ \\
				$\bar{\sigma}_d$ & $0.1m$ &
				$\bar{\sigma}_v$ & $0.25m/s$ \\  
				$\sigma^2$ & $0.15$ &
				$\sigma^2_C$ & $1$ \\		
				$G$	& $10$ & $\alpha_\text{ref}$ &  $1$ \\							
				\hline 
		\end{tabularx}}
	\end{threeparttable}
\end{table}

\section{Simulations}
In this section, we provide numerical simulation results to verify the proposed V2I beam tracking schemes. 
If not otherwise specified, the simulative parameters follows Table.~\ref{tab2}. Beyond that, we assume that the coordinate of RSU is $(0m,0m)$, and the initial position of the vehicle in the interested region is $(60m,20m)$. In simulations, we model the vehicle with resolved scatterers uniformly distributed in the spatial geometry shape.
The RCS of slowly fluctuated scatterers $\varepsilon_{k,n}$ are generated by a complex Gaussian distribution with zero mean and unit variance.
Besides, the vehicle moves in the direction of the negative $x$-axis.
All results are averaged by $500$ runs.

\subsection{Performance Evaluation of ISAC-DB Method}
\begin{figure}[!htb]
	\begin{tabular}{cc}
		\begin{minipage}[t]{0.48\linewidth}
			\includegraphics[height=6cm]{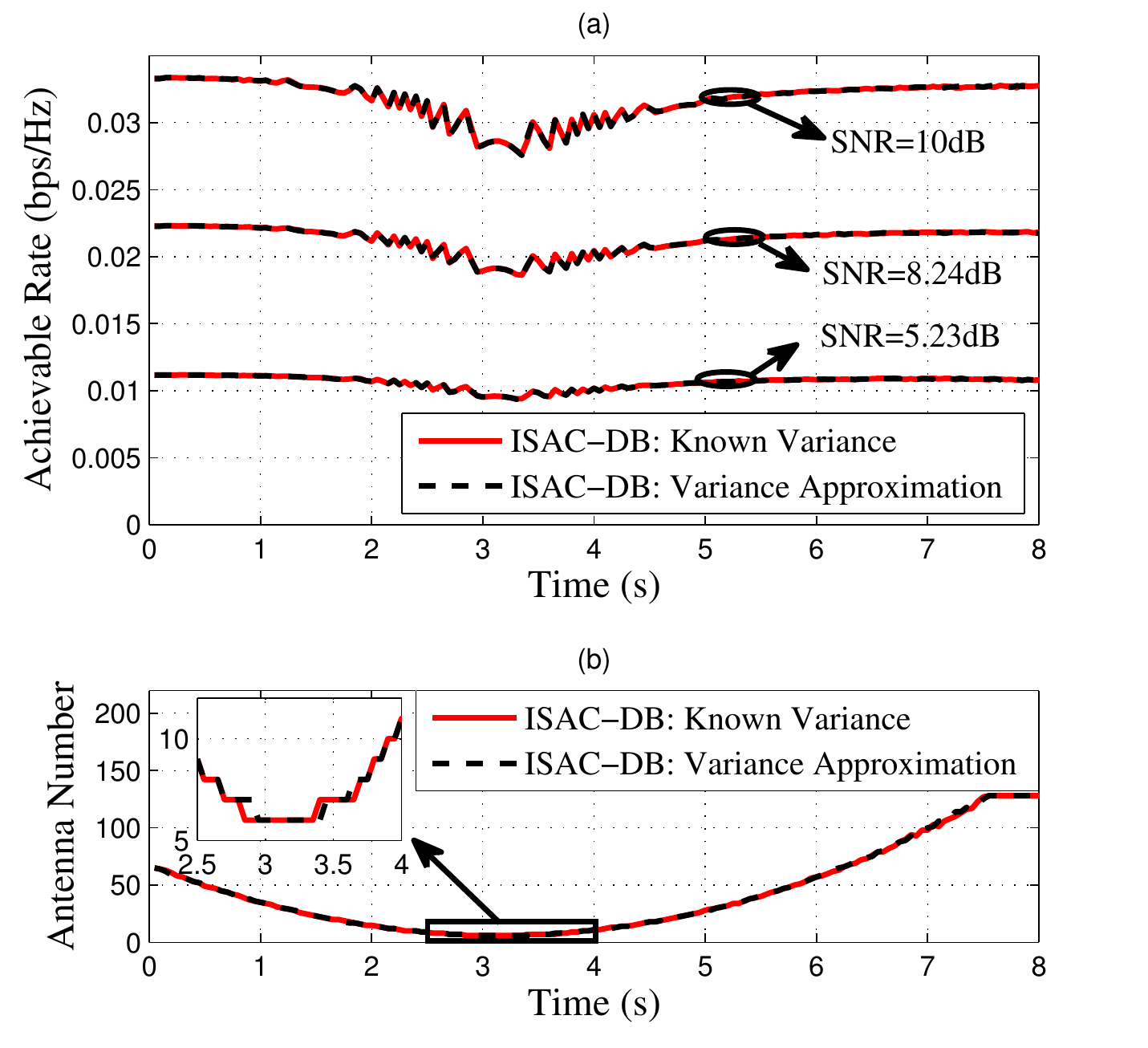}\\
			\caption{(a) Achievable rate versus time with known variances and approximated variances, respectively; (b) Antenna number $N_{t,n}$ versus time.}\label{fig40}
		\end{minipage}
		\begin{minipage}[t]{0.48\linewidth}
			\includegraphics[height=6cm]{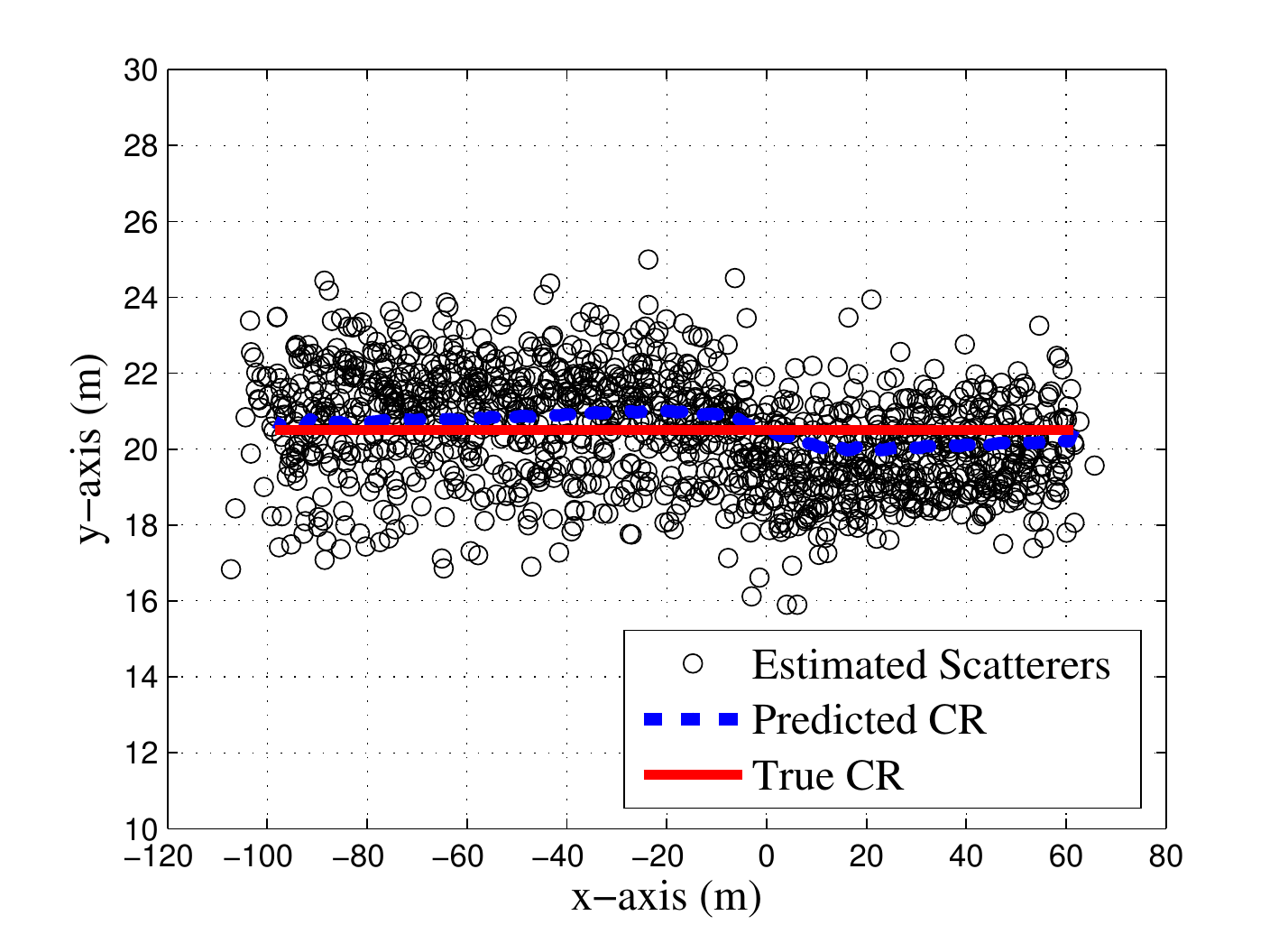}\\
			\caption{Trajectory tracking in a 2D plane.}\label{fig7}
		\end{minipage}
	\end{tabular}
\end{figure}

The achievable rates of ISAC-DB with known variances and approximated variances are provided in Fig. \ref{fig40}(a) with different transmit SNR defined as $\frac{p_n}{\sigma^2}$.
Three transmit SNR values are $\text{SNR}=5.23\text{dB}$ $(p_n=0.5, \sigma^2=0.15)$,  $\text{SNR}=8.24\text{dB}$ $(p_n=1, \sigma^2=0.15)$ and  $\text{SNR}=10\text{dB}$ $(p_n=1.5, \sigma^2=0.15)$, respectively.
Here, key findings emerge: 1) The achievable rate increases with the larger transmit SNR;
2) Approximated variances have a tiny effect in communication performance of ISAC-DB relative to the real variances, which verifies the effectiveness of variance approximations; 3) The achievable rate $R_n$ deteriorates when the vehicle approaches the RSU, which indicates that the receive SNR (i.e., $\text{SNR}^C_n$ defined in (\ref{equ22})) reduces.
For the fixed beamwidth and the transmit signal power in \cite{liu2020TWC}, it is foreseen that $R_n$ increases when the vehicle approaches the RSU due to the stronger LoS channel coefficient corresponding to the higher $\text{SNR}^C_n$. 
We also provide the numerical result of $N_{t,n}$ in Fig. \ref{fig40}(b) when $\text{SNR}=8.24\text{dB}$ with known variance and approximated variance, respectively. Interestingly, when the vehicle approaches the RSU, $N_{t,n}$ decreases for covering the entire vehicle and ISAC-DB thus achieves the lower achievable rate, which is different from the results in \cite{liu2020TWC}.
This indicates that the power attenuation caused by the smaller $N_{t,n}$ plays a more dominated role than the approaching range in affecting $\text{SNR}^C_n$. In addition, when the vehicle drives away nearby $t=8s$, the antenna number calculated by (\ref{equ25}) is fixed at $N_{t,\text{max}} = 128$, which attains the limitation of system configuration at this moment.

\begin{figure}[!htb]
	\begin{tabular}{cc}
		\begin{minipage}[t]{0.48\linewidth}
			\includegraphics[height=6cm]{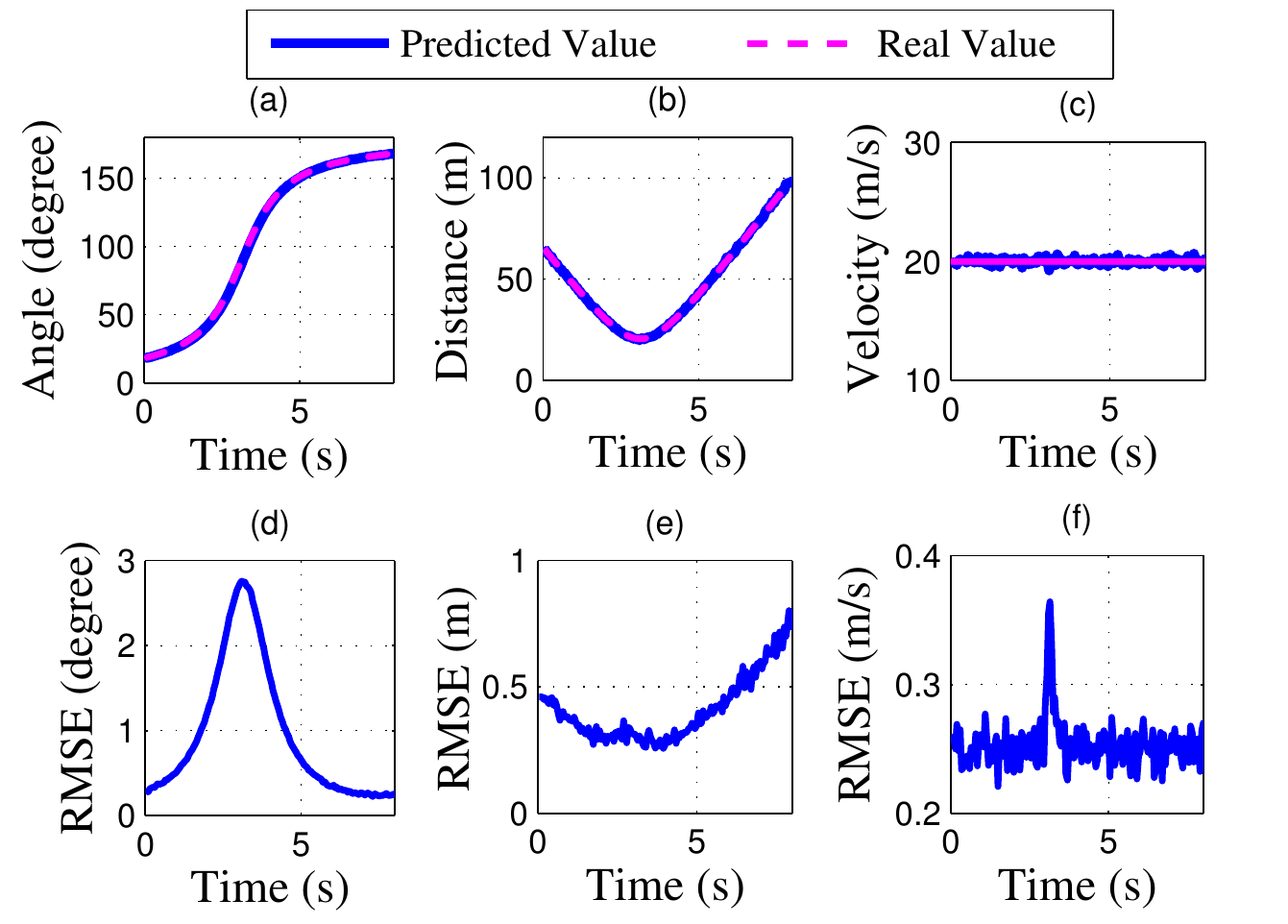}\\
			\caption{Tracking performances of the CR with ISAC-DB method. (a) Predicted angle and real angle; (b) Predicted distance and real distance; (c) Predicted velocity and real velocity; (d) RMSE of predicted angle; (e) RMSE of predicted distance; (f) RMSE of predicted velocity.}\label{fig8}
		\end{minipage}
		\begin{minipage}[t]{0.48\linewidth}
			\includegraphics[height=6cm]{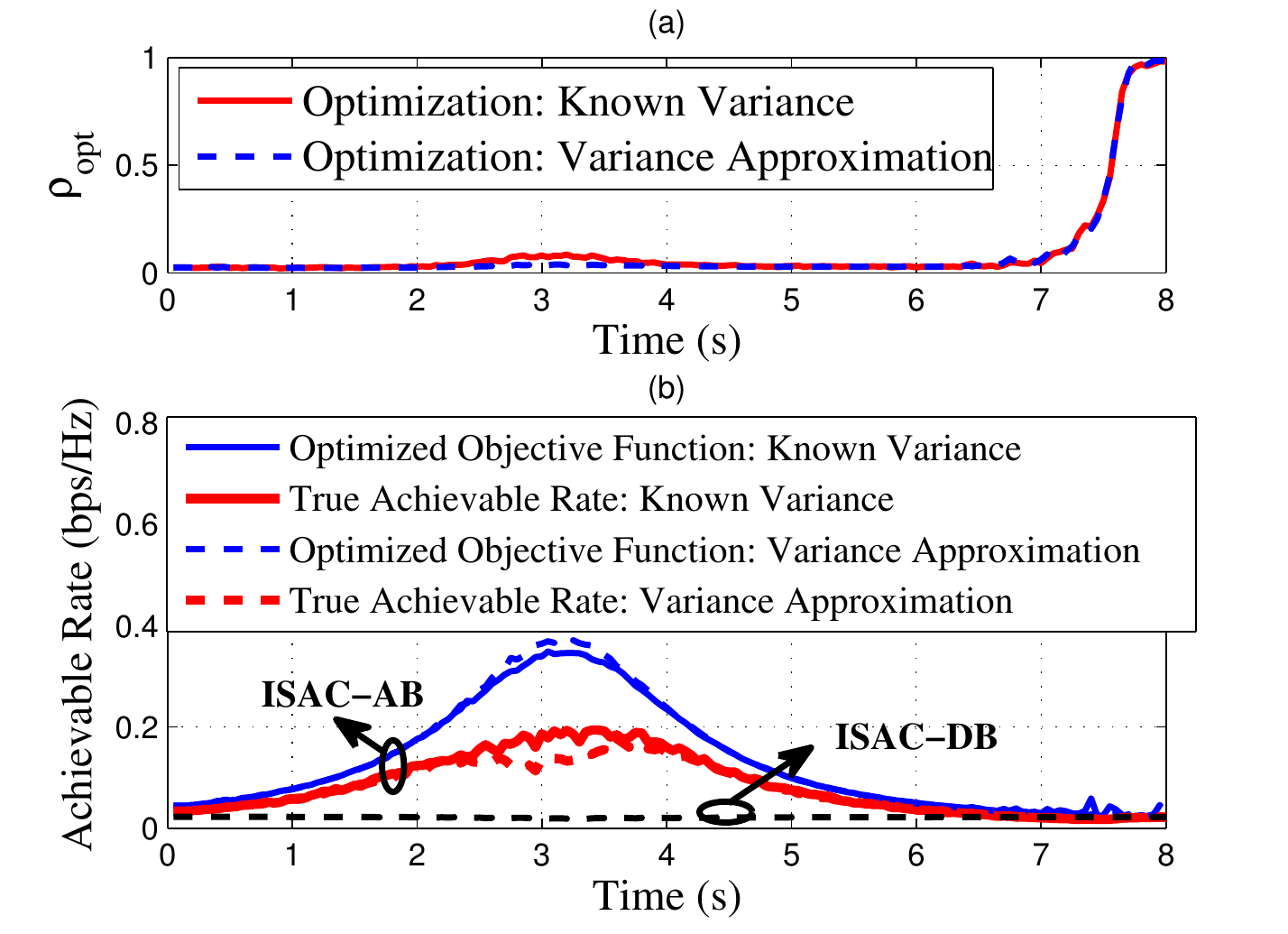}\\
			\caption{(a) Optimized $\rho_\text{opt}$ versus time with known variances and approximated variances, respectively; (b) Achievable rate versus time instant of ISAC-DB, together with ISAC-AB in four cases: optimized objective function $R^\text{obj,opt}_n$ with known variances and approximated variances, and true achievable rate $R^\text{opt}_n$ with known variances and approximated variances, respectively.}\label{fig9}
		\end{minipage}
	\end{tabular}
\end{figure}


Trajectory tracking results in a 2D plane with ISAC-DB are provided in Fig. \ref{fig7}, which include trajectories of the predicted CR and the real CR, together with the resolved scatterers at each time instant. In general, the two trajectories coincide well. When the vehicle approaches the RSU, the prediction error increases due to the decline of array gain and thereby the trajectory gap becomes relatively larger. Accordingly, predicted values and real values of angle, distance and velocity are shown in Fig. \ref{fig8}. To evaluate the prediction accuracy, the root mean square error (RMSE) curves of angle, distance and velocity are also illustrated in Fig. \ref{fig8}. It is found that the RMSE of angle increases when the vehicle approaches the RSU, in terms of the large angular variation. As for the RMSE of distance, there is no such a phenomenon because the distance variation is small at this moment. Besides, there is a peak in the RMSE of velocity, on account of the sufficiently small radial velocity measured when the RSU is approaching, leading to a relatively large prediction error. Overall, the prediction procedure exhibits a high accuracy.

\subsection{Performance Evaluation of ISAC-AB Method}
Now, we concentrate on the communication performance of the advanced ISAC-AB method with known and approximated variances, respectively. By solving (\ref{equ48}), $\rho_\text{opt}$ is obtained and shown in Fig. \ref{fig9}(a). When the vehicle approaches the RSU, the dynamic beam is with smaller array gain. Thus, to guarantee the prediction accuracy, the optimization allocates longer duration for ISAC-DB method to achieve higher matched-filtering gain.
It is also demonstrated that $\rho_\text{opt} \approx 1$ in the near region of the epoch $t=8s$ in Fig. \ref{fig9}(a). This is because when the vehicle drives away, the antenna number calculated by (\ref{equ25}) is fixed at $N_{t,\text{max}} = 128$. Then the first part and the second part of ISAC-AB are with the same antenna number.
Hence, $\rho_\text{opt} = 1$ is attained and the matched-filtering gain can not be added anymore, so that the beam tracking performance gradually deteriorates. That is why small fluctuations appear nearby the region of $t=8s$.


%
%
%

In Fig. \ref{fig9}(b), it turns out that ISAC-AB significantly outperforms ISAC-DB. As stated above, the achievable rate of ISAC-DB degrades when approaching the RSU due to the loss of array gain. However, ISAC-AB is able to perfectly overcome such a dilemma since its rate is principally contributed by the second part with a narrow beam of the high array gain. Hence, when the vehicle is approaching, the achievable rate improves in terms of the shorter distance and the fixed array gain. 

Besides, we also provide the comparison among four cases with ISAC-AB in fig. \ref{fig9}(b). Firstly, we consider the case in which the variances of $\sigma^2_{\phi}$, $\sigma^2_{d}$ and $\sigma^2_{v}$ are known.
While in Fig. \ref{fig9}(b), the true achievable rate $R^\text{opt}_n$ calculated by (\ref{equ51}) has a certain gap to the optimized objective function $R^\text{obj,opt}_n$ calculated by (\ref{equ50}). In fact, $R^\text{obj,opt}_n$ is just utilized for optimization and can not represent the real achievable rate since approximations  $\vert \mathbf{a}^H(\phi_n)\mathbf{a}(\widehat{\phi}_{n|n-1}) \vert = 1$ and $\vert \bar{\mathbf{a}}^H(\phi_n)\bar{\mathbf{a}}(\widehat{\phi}_n) \vert = 1$ are used for simplifications. This yields the performance distortion to the real value. A special case occurs when measurement variances are small enough, so that the predicted $\widehat{\phi}_{n|n-1}$ and the estimated $\widehat{\phi}_n$ equal to the true $\phi_n$, and the $P_\text{A}$ of the narrow beam equals to $1$. In that way, (\ref{equ50}) and (\ref{equ51}) are equivalent.

\begin{figure}[!htb]
	\begin{tabular}{cc}
		\begin{minipage}[t]{0.48\linewidth}
			\includegraphics[height=6cm]{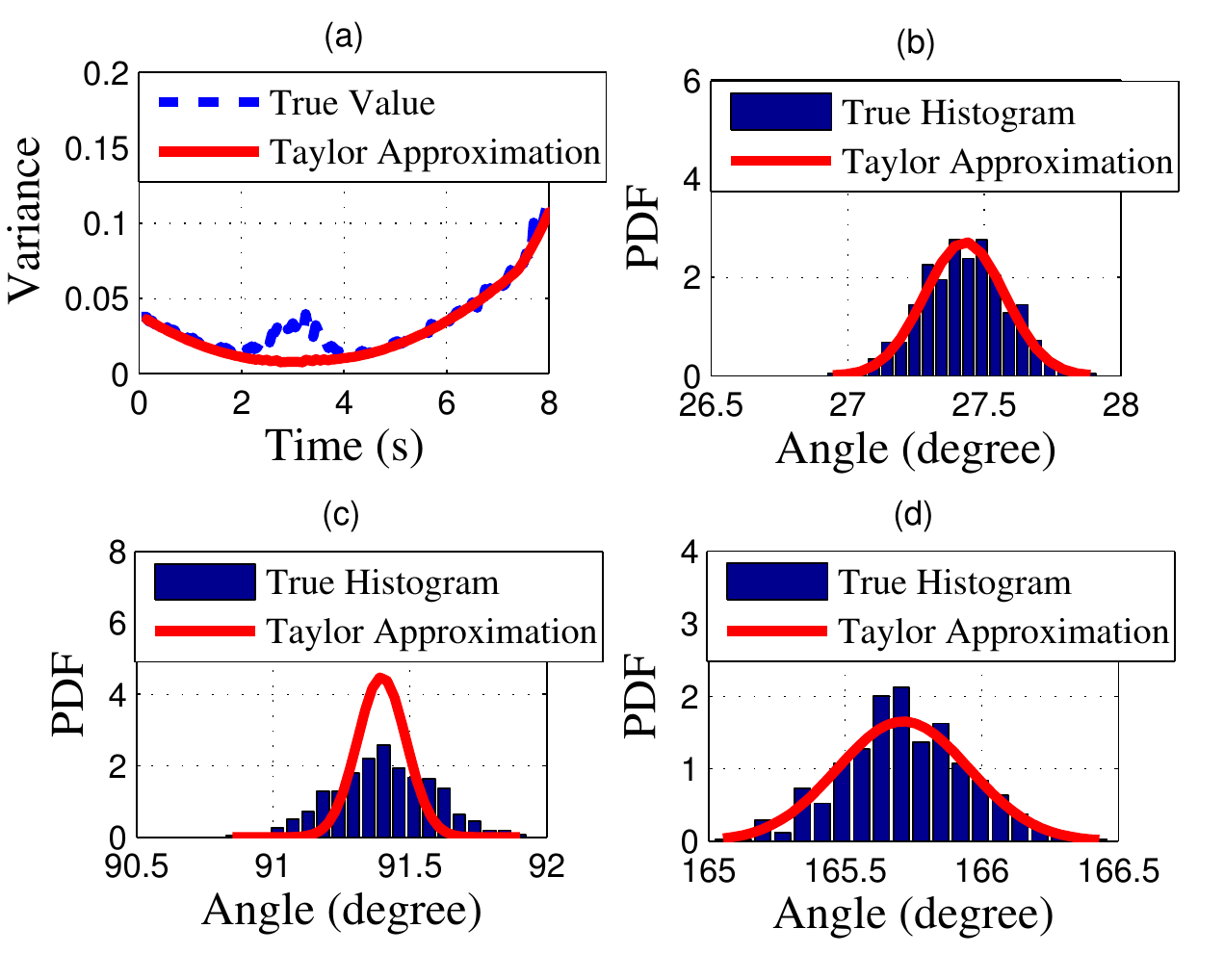}\\
			\caption{True angle variance and approximated counterpart with the first-order Taylor expansion. (a) Variance versus time; (b) PDF versus angle at $t=1s$; (c) PDF versus angle at $t=3s$; (d) PDF versus angle at $t=7s$.}\label{fig10}
		\end{minipage}
		\begin{minipage}[t]{0.48\linewidth}
			\includegraphics[height=6cm]{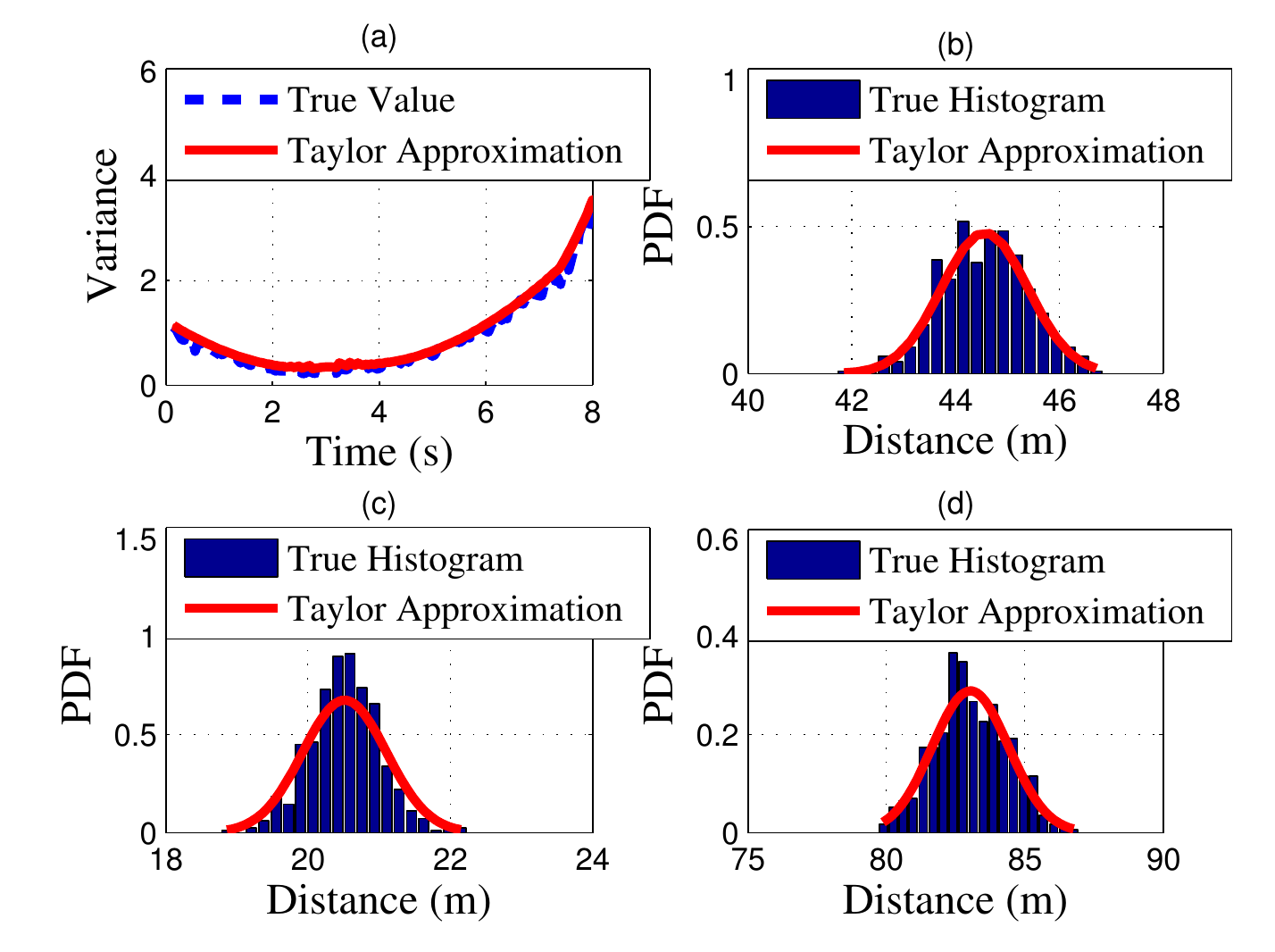}\\
			\caption{True distance variance and approximated counterpart with the first-order Taylor expansion. (a) Variance versus time; (b) PDF versus angle at $t=1s$; (c) PDF versus angle at $t=3s$; (d) PDF versus angle at $t=7s$.}\label{fig11}
		\end{minipage}
	\end{tabular}
\end{figure}

Now, we consider the case in which the variances of $\sigma^2_{\phi}$, $\sigma^2_{d}$ and $\sigma^2_{v}$ are unknown, and validate the availability of the developed variance approximations. 
It is found that the approximation accuracy of $\sigma^2_v$ has a tiny effect on the beam tracking in simulations. Here, regardless of the approximated $\sigma^2_v$ obtained by (\ref{equ67}), we only investigate the the feasibility of the first-order Taylor expansions on $\sigma^2_{\phi}$ and $\sigma^2_{d}$. Note that the true variaces  of angle and distance at the $n$th epoch are calculated as $\sigma^2_{\phi} = E[(\widehat{\phi}_n-\phi_n)^2]$ and $\sigma^2_{d} = E[(\widehat{d}_n-d_n)^2]$. In Fig. \ref{fig10}(a), the true variance $\sigma^2_{\phi}$ becomes smaller when the vehicle approaches the RSU, and raises when it drives away. However, at surroundings of the epoch $t=3s$ corresponding to $\phi_n \approx \pi/2$, the angles of scatterers changes too fast in the high mobility scenarios, which causes the spikes for the estimation of $\phi_n$.
Meanwhile, it is found that the similar approximation of $\sigma^2_d$ in Fig. \ref{fig11}(a) performs better when $\phi_n \approx \pi/2$, since the distances of scatterers changes more slowly at this moment.
Fortunately, the spikes of angle would not seriously affect the achievable rate, despite that some small fluctuations appear in the corresponding epoches. As shown in Fig. \ref{fig9}(b), the true achievable rate with approximated variance has a slight performance loss relative to the case with known variance, which is caused by the approximation errors. Note however that the optimized objective function with approximated variances exceeds the optimized objective function with true variances when $\phi_n \approx \pi/2$, since the approximated $\sigma^2_{\phi}$ at this time is slightly smaller than the true value. Therefore, the smaller $\sigma^2_{\phi}$ leads to the smaller $\rho_\text{opt}$, as shown in Fig. \ref{fig9}(a).

Figs. \ref{fig10}(b)-(d) and Figs. \ref{fig11}(b)-(d) also provide the histograms of the measured $\widehat{\phi}_n$ and $\widehat{d}_n$ at epoches $t=1s$, $t=3s$ and $t=7s$, respectively. Overall, the probability density functions (PDFs) of Gaussian fittings with approximated variances can well match the histograms. This again verifies that the feasibility of Gaussian assumptions and the effectiveness of variance approximations.


\subsection{Performance Comparison Among the Proposed Schemes and State-of-The-Art Methods}
We examine the superiorities of the proposed ISAC-DB and ISAC-AB by comparing with the following beam tracking algorithms:
\begin{itemize}
	\item[$\bullet$] The auxiliary beam pair (ABP) algorithm, where a pair of training beams are transmitted within each block \cite{zhu2017auxiliary};
	\item[$\bullet$] EKF-based beam tracking algorithm for point target (denoted as `` EKF-Point Target'' in the following) \cite{liu2020TWC}. 
\end{itemize}
Note that ABP algorithm belongs to the class of the communication-only beam tracking which needs an additional uplink feedback, of which the overhead in front of each data block is much higher than the ISAC-based beam tracking. For the EKF-Point Target, as we highlight before, it is unable to precisely track the CR when the vehicle is extended in range and angle domains. For a fair comparison, we randomly select a scatterer of the vehicle (not in the position of CR) and track it. For the RSU, it treats this scatterer as the CR since the target is considered as point-like for EKF-Point Target. Note that both the two schemes are parameterized with a fixed antenna number as $N_t=128$. For ABP, the half searching range of the beamforming region for the transmitter defined in \cite{zhu2017auxiliary} is set as $\frac{\pi}{32}$ without loss of generality.

\begin{figure}[!htb]
	\begin{tabular}{cc}
		\begin{minipage}[t]{0.48\linewidth}
			\includegraphics[width=3in]{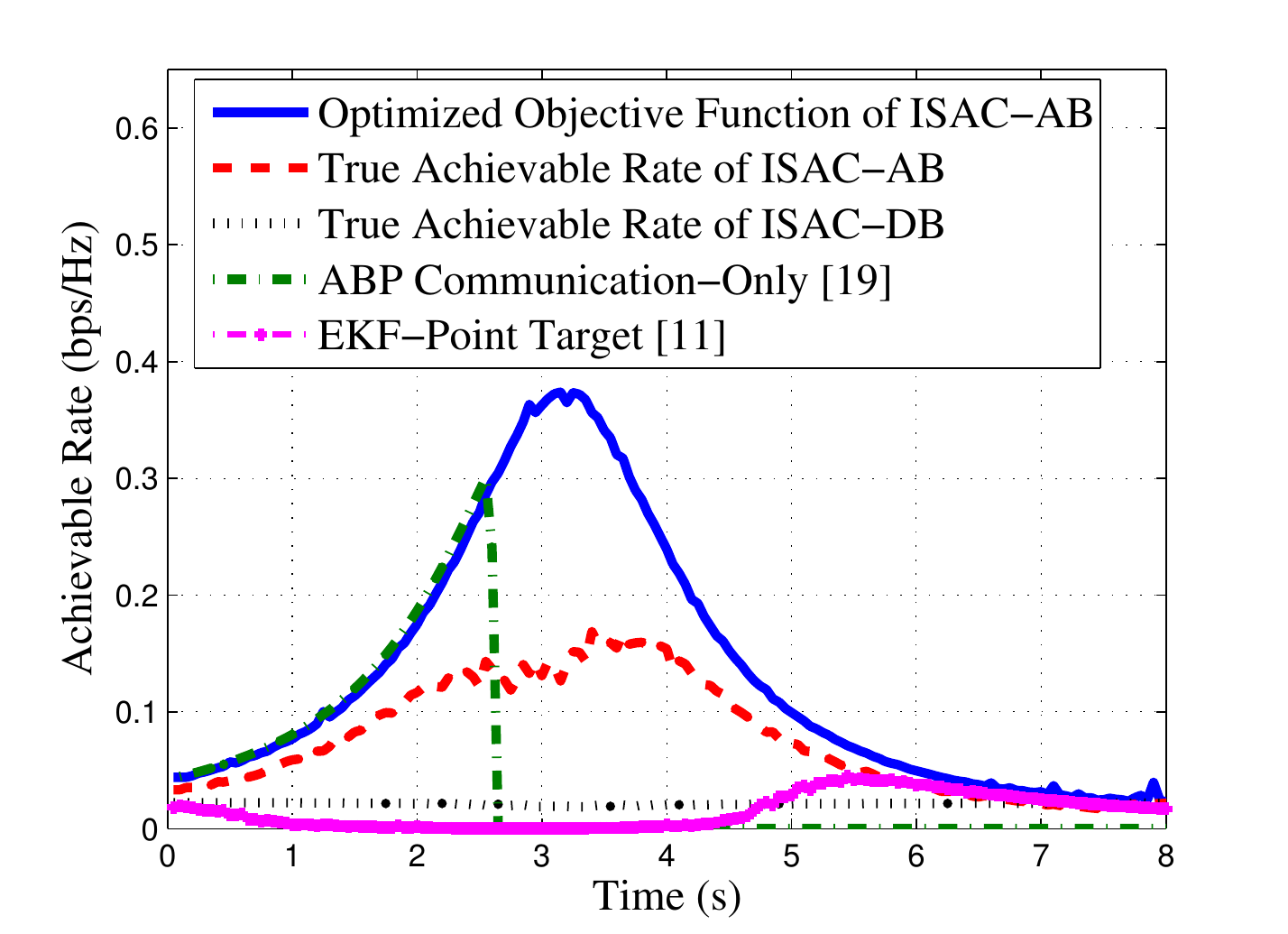}
	\caption{Achievable rate comparison among ISAC-AB, ISAC-DB, ABP, and EKF-Point Target methods.}\label{fig12}
		\end{minipage}
		\begin{minipage}[t]{0.48\linewidth}
			\includegraphics[width=3in]{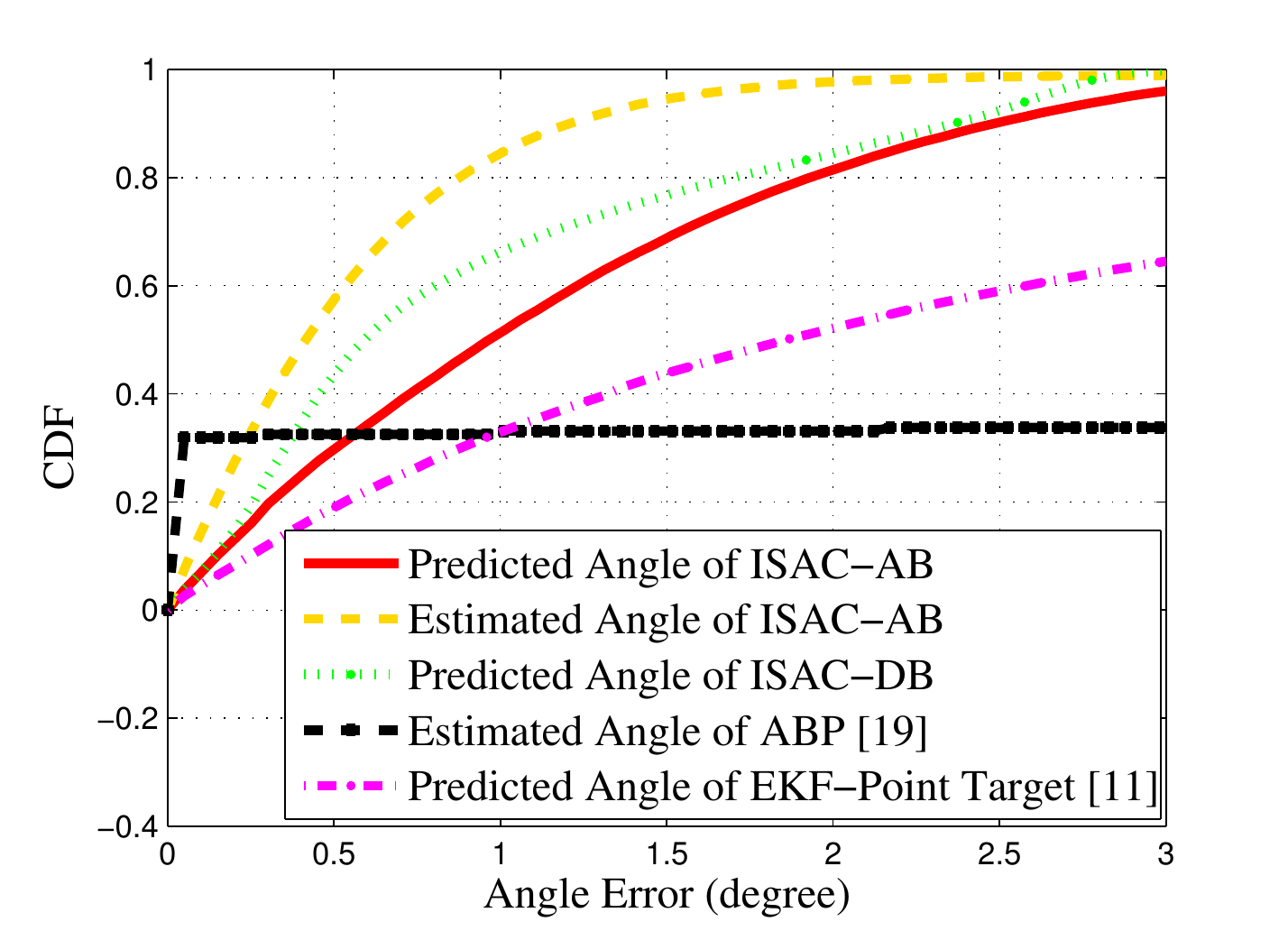}
	\caption{CDF comparison of overall angle tracking among ISAC-AB, ISAC-DB, ABP, and EKF-Point Target methods.}\label{fig14}
		\end{minipage}
	\end{tabular}
\end{figure}

%
%
%
%
%
%
\begin{figure}[!htb]
	\centering
	\includegraphics[width=3in]{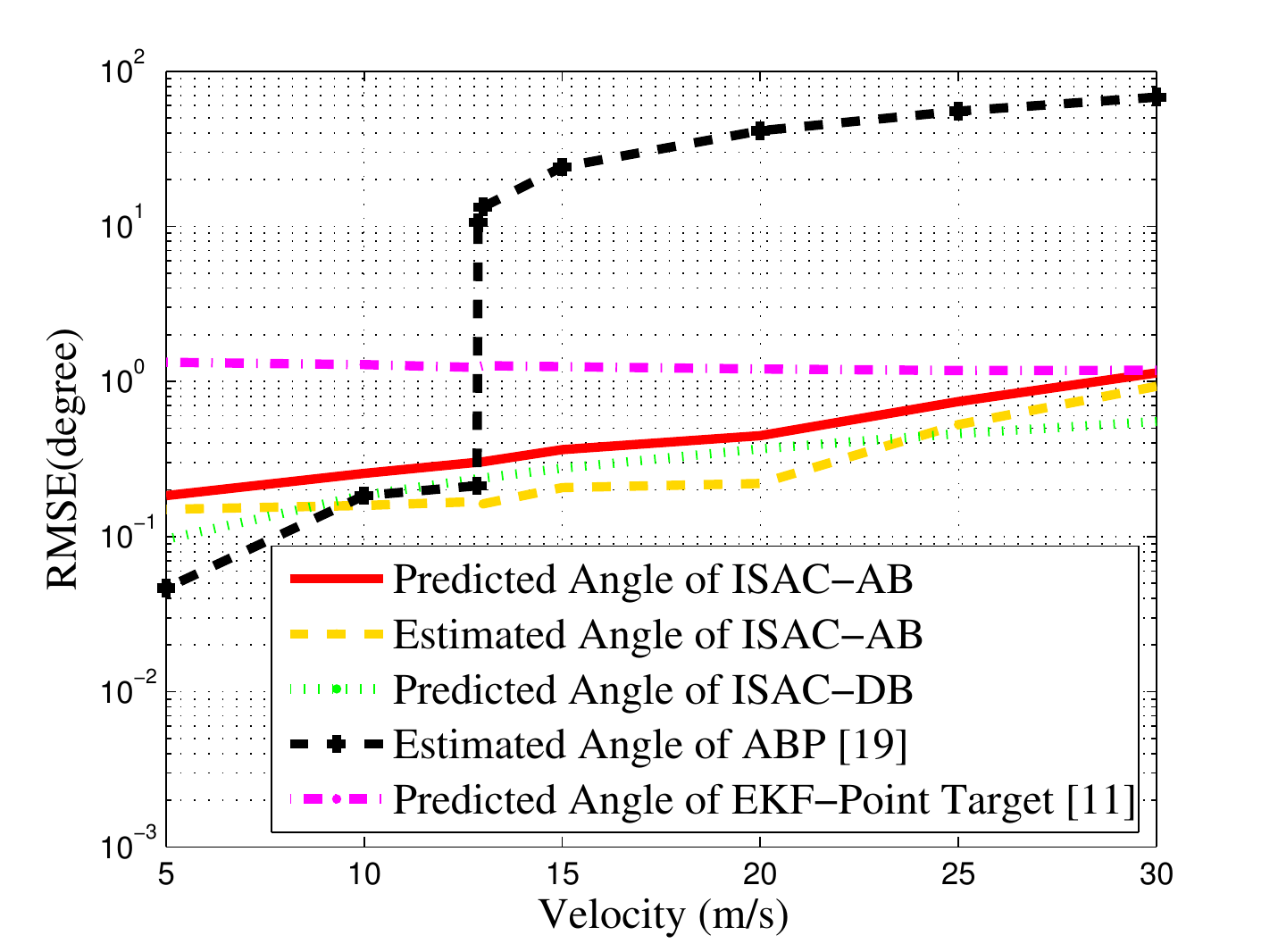}\\
	\caption{RMSE comparison of overall angle versus velocity among ISAC-AB, ISAC-DB, ABP, and EKF-Point Target methods.}\label{fig17}
\end{figure}

The comparative results of these schemes are provided in Fig. \ref{fig12}. It is interesting to see that ISAC-AB achieves better performance than EKF-Point Target, while is inferior to ABP in the region of $t \in [0s, 2.6s]$. To account for this, we emphasize that ABP is a communication-only beam tracking algorithm with a fixed antenna number $N_t=128$ corresponding to a narrow beam with the highest array gain. Hence, it surpasses ISAC-AB only if the tracking is successful, since the achievable rate of ISAC-AB is a combination of both the wide beam and the narrow beam. However, when the vehicle approaches the RSU, the angle changes faster, and the angular variation is sufficiently large which is not likely to fall into the searching range of the beamforming region as $\frac{\pi}{16}$. As such, the vehicle trajectory is lost and the beam tracking fails. As for the EKF-Point Target, it is much inferior to ISAC-AB due to the beam misalignment. In fact, when the vehicle approaches the RSU, the narrow beam with fixed $N_t=128$ antennas is more unlikely to simultaneously cover both the tracked scatterer and CR, arising a drastic beam misalignment.

For the sake of a deeper perception, we further show the overall tracking performance via the cumulative distribution function (CDF) of angle prediction and/or estimation errors in Fig. \ref{fig14}. The results explicitly support the analysis above.


To show the superiority of the proposed methods in high-mobility scenarios, we firstly demonstrate the overall RMSE results of angle versus the velocity in Fig.\ref{fig17}. The simulative velocities are set as $30m/s$, $25m/s$, $20m/s$, $15m/s$, $13m/s$, $12.875m/s$, $12.85m/s$, $10m/s$ and $5m/s$, respectively. The vehicle moves in the same trajectory of length $160m$. In general, ISAC-AB, ISAC-DB and ABP methods achieve worse RMSE tracking results with an increased velocity, because of a nonnegligible movement in each epoch interval $\Delta T$. For ISAC-AB and ISAC-DB methods, a lower velocity helps the CR to align the beam in $\Delta T$. As for ABP, it is sensitive to the velocity. Especially when the velocity is larger than $12.875m/s$, ABP method rapidly breaks down since the angular variation is too large to fall into the searching range, which again verifies the superiority of ISAC-AB method over the conventional communication-only beam tracking methods in high mobility scenarios. 
Interestingly, EKF-Point Target method behaves slightly better with an increased velocity. Since it always tracks a selected scatterer, the CR may be closer to the narrow beam aligned to the scatterer, considering the movement in $\Delta T$. Therefore, when the vehicle moves faster, the CR has a higher probability to lie in the narrow beam if the scatterer is ahead of the CR in the direction of movement.

%

\begin{figure}[!htb]
	\begin{tabular}{cc}
		\begin{minipage}[t]{0.48\linewidth}
			\includegraphics[width = 3in]{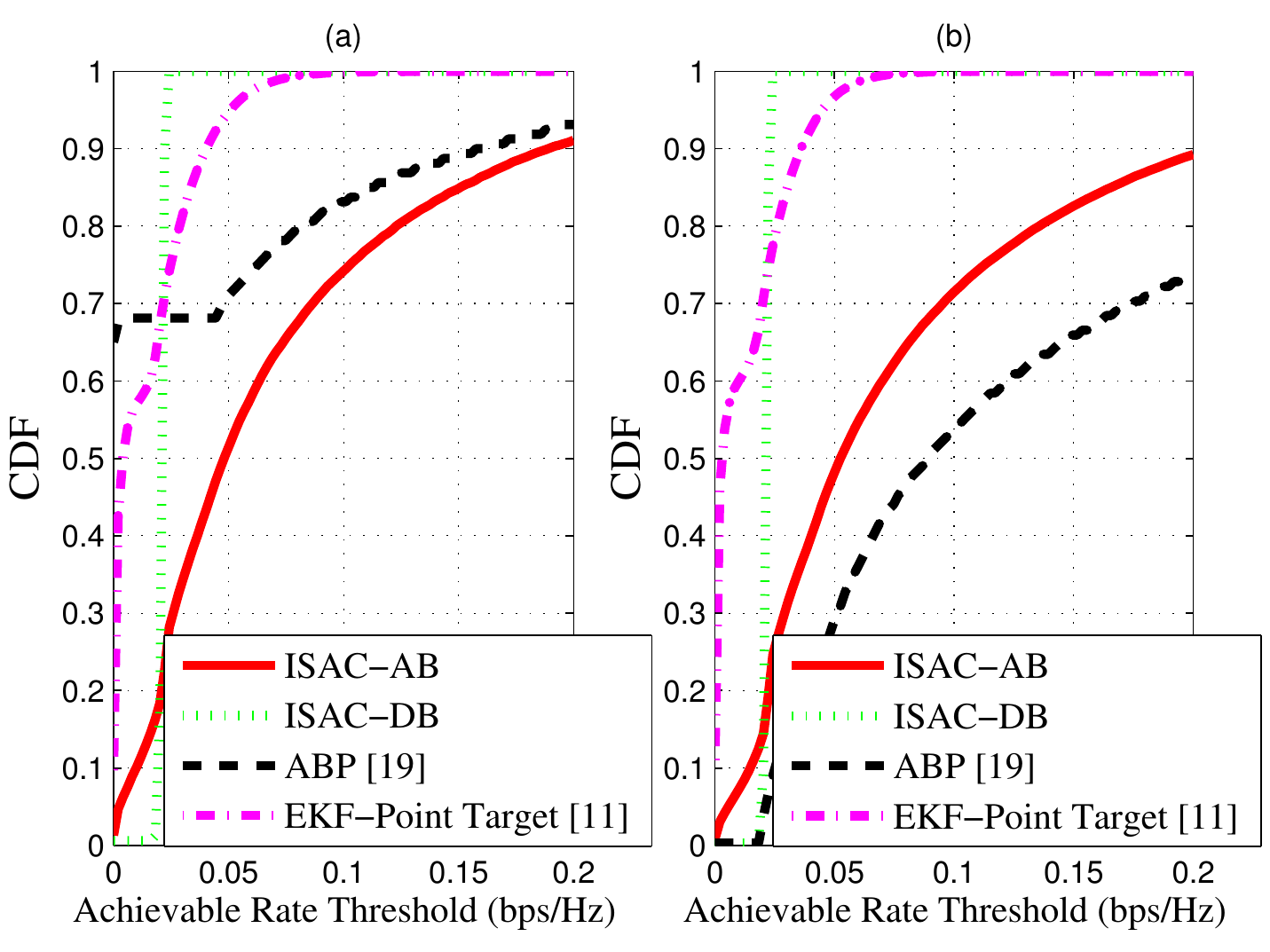}
	\caption{CDF versus achievable rate threshold among ISAC-AB, ISAC-DB, ABP, and EKF-Point Target methods. (a) $v=20m/s$; (b) $v=10m/s$.}\label{fig160}
		\end{minipage}
		\begin{minipage}[t]{0.48\linewidth}
			\includegraphics[width = 3in]{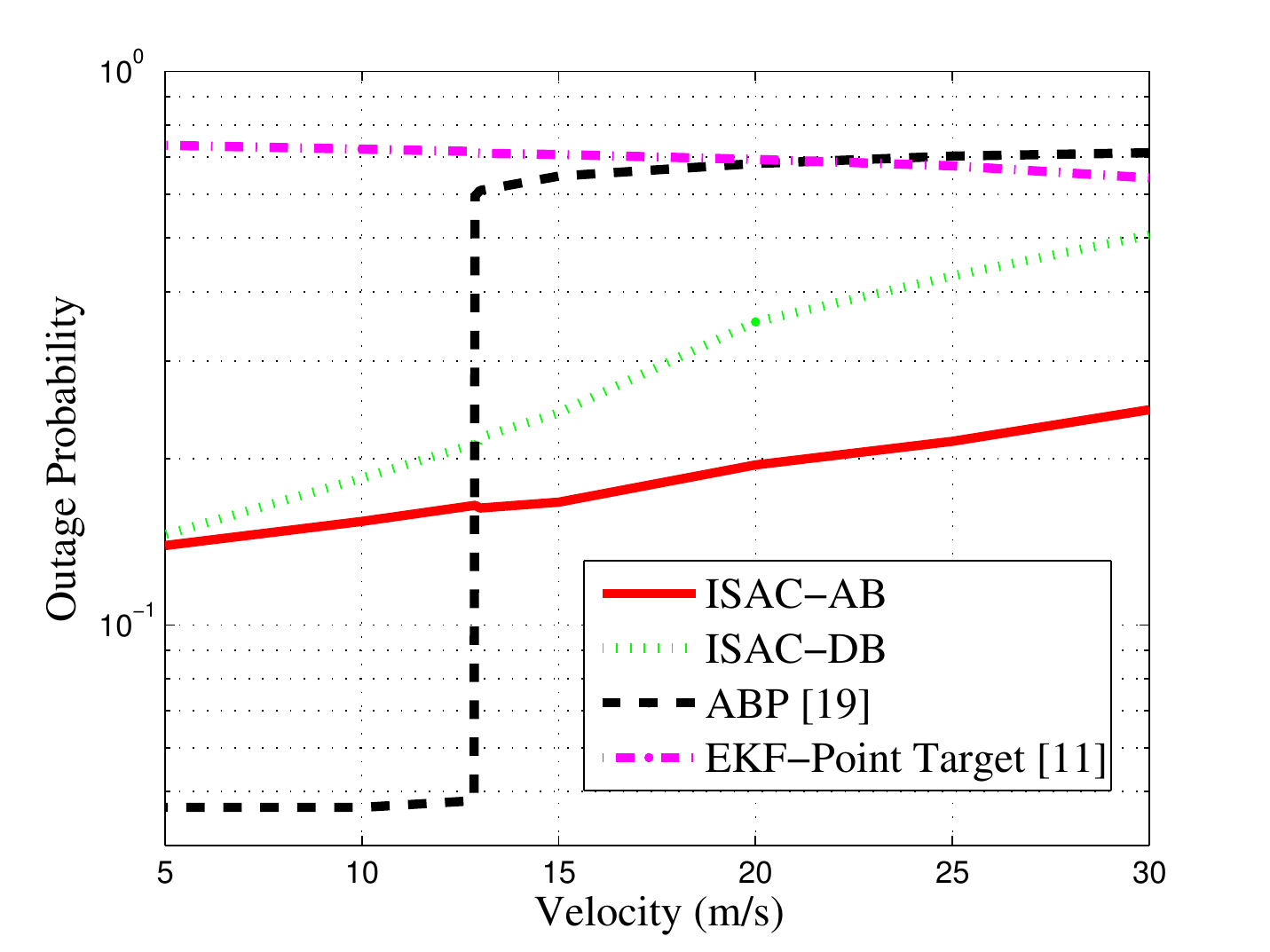}
	\caption{Outage probability of overall achievable rate versus velocity among ISAC-AB, ISAC-DB, ABP, and EKF-Point Target methods.}\label{fig16}
		\end{minipage}
	\end{tabular}
\end{figure}

Next, we define the outage probability of overall achievable rate, which is given as
\begin{equation}
	\begin{aligned}
		P_\text{OUT} = \frac{1}{T/\Delta T}\sum\nolimits^{T/\Delta T}_{n=1}\text{Prob}\left\{ R_n \leq \gamma \right\},
	\end{aligned}
\end{equation}
where $\gamma$ denotes the achievable rate threshold.
In Fig. \ref{fig160}, we first demonstrate the CDF curves versus the achievable rate threshold, in cases of $v=20m/s$ and $v=10m/s$, respectively. Overall, when $v=20m/s$, ISAC-AB method behaves better than ISAC-DB, ABP and EKF-Point Target methods in achieving higher rates. However, when $v=10m/s$, ABP method outperforms other methods in general, benefiting from the low-mobility that the angular variation is small enough to fall into the searching range. In this case, ABP method is even superior to ISAC-AB method since the beam is accurately aligned to the CR and the narrow beam with $128$-antenna array is used in each entire block length $\Delta T$. 

For a further insight of the velocity impact, the outage probability curves versus different velocities are provided in Fig. \ref{fig16}, where the achievable rate threshold is set as $\gamma=0.02$bps/Hz\footnote{For example, the mmWave signaling with a bandwidth as $500$MHz is qualified to support the data transmission with $10$Mbps data rate.}. 
It can be concluded that the proposed ISAC-AB and ISAC-DB methods, along with ABP method, achieve better communication transmission when the velocity is lower. Likewise, it is found that ABP method is sensitive to the velocity, as stated before. Overall, ISAC-AB method outperforms other methods in high-mobility scenarios, in terms of the outage probability performance. 

\section{Conclusion}
In this article, we have proposed novel schemes in regard to sensing-assisted mMIMO beam tracking for the extended target vehicle. Exploiting such an ISAC approach provides significant gains in terms of both sensing and communications.
We have proposed EKF beam prediction approaches that exploit varying beamwidths. Our approaches, thanks to the extended target modeling, offer significant enhancements to state-of-the-art ISAC solutions by optimizing the beam tracking to focus on the CR.
Finally, numerical results have verified the feasibility and the effectiveness of the proposed optimization scheme, which has remarkable superiorities over the state-of-the-art beam tracking approaches.

\appendix[Approximations of measurement variances in (\ref{equ18})]

Note that all $\widehat{\theta}_{k,n}$, $\widehat{d}_{k,n}$ and $\widehat{\mu}_{k,n}$ are Gaussian variables.
However, due to the non-linearity of (\ref{equ18}), $\widehat{\phi}_n$, $\widehat{d}_n$ and $\widehat{v}_n$ are not Gaussian, thus $\sigma^2_{\phi}$, $\sigma^2_d$ and $\sigma^2_v$ are intractable. To resolve this problem, we resort to the first-order Taylor expansion for approximations of $\sigma^2_{\phi}$ and $\sigma^2_d$, together with a coarse approximation of $\sigma^2_v$.

To be specific, firstly we denote $g(\mathbf{q}_n) \triangleq \phi_n$ where $\mathbf{q}_n = [\theta_{1,n},\cdots,\theta_{K,n},d_{1,n},\cdots,d_{K,n}]^T$.
Then with the help of first-order Taylor expansion at the estimation $\widehat{\mathbf{q}}_n$, we have
\begin{equation}\label{equ53}
	\begin{aligned}
		g(\mathbf{q}_n) \approx g(\widehat{\mathbf{q}}_n) + \frac{\partial g}{\partial \mathbf{q}^T_n}\bigg\arrowvert _{\mathbf{q}_n=\widehat{\mathbf{q}}_n} (\mathbf{q}_n - \widehat{\mathbf{q}}_n) 
		= g(\widehat{\mathbf{q}}_n) + \mathbf{G}_n (\mathbf{q}_n - \widehat{\mathbf{q}}_n),
	\end{aligned}
\end{equation}
where $\widehat{\phi}_n=g(\widehat{\mathbf{q}}_n)$, and $\mathbf{G}_n = \frac{\partial g}{\partial \mathbf{q}^T_n}\bigg\arrowvert_{\mathbf{q}_n=\widehat{\mathbf{q}}_n} \in \mathbb{R}^{1\times 2K}$ 
denote the Jacobian matrix of the angle with respect to $\mathbf{q}_n$.
Here, it is a row vector with the form as
\begin{equation}\label{equ54}
	\begin{aligned}
		\mathbf{G}_n = \Bigg[ \frac{\partial g}{\partial \theta_{1,n}},\cdots, \frac{\partial g}{\partial \theta_{K,n}}, \frac{\partial g}{\partial d_{1,n}},\cdots,  \frac{\partial g}{\partial d_{K,n}} \Bigg]_{\mathbf{q}_n=\widehat{\mathbf{q}}_n}.
	\end{aligned}
\end{equation}
Now, for brief notations, we denote 
\begin{align}
	\Delta X = \sum^K_{k=1} {d_{k,n}\cos \theta_{k,n}} + K\Delta x, \
	\Delta Y = \sum^K_{k=1} {d_{k,n}\sin \theta_{k,n}} + K\Delta y.
\end{align}
Hence, the elements of (\ref{equ54}) are derived as
\begin{align}					
	\frac{\partial g}{\partial \theta_{k',n}} = \frac{ d_{k',n}\cos \theta_{k',n} \Delta X + d_{k',n} \sin \theta_{k',n} \Delta Y } {\Delta X^2 + \Delta Y^2}, \ 
	\frac{\partial g}{\partial d_{k',n}} = \frac{ \sin \theta_{k',n} \Delta X  - \cos \theta_{k',n} \Delta Y } {\Delta X^2  + \Delta Y^2}.
\end{align}
Next, a reformulation of (\ref{equ53}) is thus obtained as
\begin{equation}
	\begin{aligned}
		g(\widehat{\mathbf{q}}_n) \approx & g(\mathbf{q}_n) + \mathbf{G}_n \left(\widehat{\mathbf{q}}_n - \mathbf{q}_n\right).
	\end{aligned}
\end{equation}
Now, given $\widehat{\mathbf{q}}_n \sim \mathcal{N}(\mathbf{q}_n,\mathbf{\Sigma}_n)$, it is able to approximate $g(\widehat{\mathbf{q}}_n) \sim \mathcal{N} \left( g(\mathbf{q}_n),  \mathbf{G}_n\mathbf{\Sigma}_n\mathbf{G}^T_n  \right)$.
Since the elements of $\mathbf{q}_n$ are independent of each other, then we have 
\begin{equation}
	\begin{aligned}
		\mathbf{\Sigma}_n = \text{diag} \big( \sigma^2_{1,n}(1),\cdots, \sigma^2_{K,n}(1),\sigma^2_{1,n}(2),\cdots,\sigma^2_{K,n}(2) \big).
	\end{aligned}
\end{equation}
Finally, the angle variance of the CR is able to be approximated as
\begin{equation}
	\begin{aligned}
		\sigma^2_{\phi} \approx \mathbf{G}_n\mathbf{\Sigma}_n\mathbf{G}^T_n.
	\end{aligned}
\end{equation}

Similarly, we denote $f(\mathbf{q}_n) \approx f(\widehat{\mathbf{q}}_n) + \mathbf{F}_n (\mathbf{q}_n - \widehat{\mathbf{q}}_n)$, where $f(\mathbf{q}_n) \triangleq d_n $, and $\mathbf{F}_n = \frac{\partial f}{\partial \mathbf{q}^T_n}\bigg\arrowvert _{\mathbf{q}_n=\widehat{\mathbf{q}}_n} \in \mathbb{R}^{1\times 2K}$ denotes the Jacobian matrix of the distance with respect to $\mathbf{q}_n$, which has a similar form as $\mathbf{G}_n$. 
Hence, the elements of $\mathbf{F}_n$ are derived as
\begin{align}			
	\frac{\partial f}{\partial \theta_{k',n}} = \frac{ -d_{k',n}\sin \theta_{k',n} \Delta X + d_{k',n} \cos \theta_{k',n} \Delta Y } {K \sqrt{\Delta X^2 + \Delta Y^2}}, \
	\frac{\partial f}{\partial d_{k',n}} = \frac{ \cos \theta_{k',n} \Delta X  + \sin \theta_{k',n} \Delta Y } {K \sqrt{\Delta X^2 + \Delta Y^2}}.
\end{align}
Finally, the distance variance of CR is able to be approximated as
\begin{equation}
	\begin{aligned}
		\sigma^2_{d} \approx \mathbf{F}_n\mathbf{\Sigma}_n\mathbf{F}^T_n.
	\end{aligned}
\end{equation}

As for the approximation of $\sigma^2_v$, it can not be conducted with the first-order Taylor expansion due to $v_n \neq \frac{c}{2f_c} \cdot \frac{\sum^{K}_{k=1}{\mu}_{k,n}\cos \left({\theta}_{k,n}\right)/\sigma^2_{k,n}(3)}{\sum^{K}_{k=1}\cos^2\left({\theta}_{k,n}\right)/\sigma^2_{k,n}(3)}$. However, we have noticed that $\widehat{v}_{n|\bm{\theta}_n} \sim \mathcal{N} \left(v_n,\left(\mathbf{A}^T(\bm{\theta}_n) \mathbf{Q}^{-1}_{\mu}\mathbf{A}(\bm{\theta}_n) \right)^{-1} \right)$, which is based on the assumption of known $\bm{\theta}_n$. In fact, simulations demonstrate that the approximated accuracy of $\sigma^2_v$ has little impact on the performance of beam tracking. Therefore, we may instead use a coarse approximation as 
\begin{equation}\label{equ67}
	\begin{aligned}
		\sigma^2_{v} \approx \left(\mathbf{A}^T(\bm{\widehat{\theta}}_n) \mathbf{Q}^{-1}_{\mu}\mathbf{A}(\bm{\widehat{\theta}}_n) \right)^{-1}.
	\end{aligned}
\end{equation}

\ifCLASSOPTIONcaptionsoff
\newpage
\fi

\bibliographystyle{IEEEtran}
\bibliography{bare_jrnl}


\end{document}